\documentclass[%
 aip,
 jmp,%
 amsmath,amssymb,
preprint,%
]{revtex4-1}

\usepackage{graphicx}
\usepackage{dcolumn}
\usepackage{bm}
\usepackage{amsthm}
\usepackage{color}

\newcommand{\T}[1]{\tilde{#1}}

\newtheorem{proposition}{Proposition}

\begin{document}

\title[Peristlatic contractions]{Optimal reservoir conditions for fluid extraction through permeable walls in the viscous limit} 

\author{G. Herschlag}
\email{gjh@math.duke.edu.}
\affiliation{Mathematics Department, Duke University.}

\author{J.-G. Liu}%
\affiliation{Mathematics Department, Duke University.}
\affiliation{Physics Department, Duke University.}

\author{A. T. Layton}%
\affiliation{Mathematics Department, Duke University.}

\date{\today}


\begin{abstract}
In biological transport mechanisms such as insect respiration and renal filtration, fluid travels along a leaky channel allowing exchange with systems exterior the the channel.  The channels in these systems may undergo peristaltic pumping which is thought to enhance the material exchange.  To date, little analytic work has been done to study the effect of pumping on material extraction across the channel walls.  In this paper, we examine a fluid extraction model in which fluid flowing through a leaky channel is exchanged with fluid in a reservoir.  The channel walls are allowed to contract and expand uniformly, simulating a pumping mechanism.  
In order to efficiently determine solutions of the model, we derive a formal power series solution for the Stokes equations in a finite channel with uniformly contracting/expanding permeable walls.  This flow has been well studied in the case of weakly permeable channel walls in which the normal velocity at the channel walls is proportional to the wall velocity.  In contrast we do not assume weakly driven flow, but flow driven by hydrostatic pressure, and we use Dacry's law to close our system for normal wall velocity.  
We use our flow solution to examine flux across the channel-reservoir barrier and demonstrate that pumping can either enhance or impede fluid extraction across channel walls.  We find that associated with each set of physical flow and pumping parameters, there are optimal reservoir conditions that maximizes the amount of material flowing from the channel into the reservoir.
\end{abstract}

\maketitle

\section{Introduction}
Flow through permeable channels at small Reynolds number occurs in a variety of physical and biological applications.  
In air flow in trachea\cite{WeisFogh1976, Komai1998} and urine in renal collecting ducts\cite{SchmidtNielsen2011}, fluid flow is altered via muscle contractions that can deform and change the cross sectional area of transport systems.  
%
%
In the instance of insect respiration, it is experimentally understood that pumping due to thoracic or dorsal ventral muscles enhances oxygen intake and carbon dioxide output \cite{Komai1998}; however, it is unclear if mechanism behind the enhanced respiration arises primarily from the increased advection leading to refreshed air and hence enhanced diffusion across the tracheal membrane, or from secondary pressure-driven flows across the tracheal membranes due to the motion at the wall.  In renal filtration a similar question persists in experimental studies and it remains controversial to what degree pumping, caused by muscle contraction along the papilla, augments the molarity of waste products and salts in urine  \cite{SchmidtNielsen2011}.

To study this problem, we consider a simple filtration model in which fluid is exchanged between a channel and a reservoir (or interstitium). 
The problem of fluid flow through permeable channels has been studied extensively over the past 60 years.  A common physical assumption is that the region exterior to the channel has a constant and fixed pressure, and fluid flow across the channel wall is driven by the hydrostatic pressure difference, typically modeled linearly with Darcy's law.  The seminal work on this problem was performed by Regirer\cite{Regirer60}, who determined an asymptotic solution for flow through a channel with low permeability and low Reynolds number.  Subsequently, the problem of flow across channels driven by Darcy's law has been expanded in a number of ways \cite{Galowin1974,Granger1988,Karode2001,Haldenwang2007,Haldenwang2011,Tilton2012,Bernales2014, Herschlag:15}.  A related problem considering prescribed cross-channel velocities in pipes and channels has also been the subject of significant investigation, beginning in 1953 with Berman \cite{Berman1953}, and this problem has also been extended in a variety of ways \cite{YuanFinkelstein1956, Terrill1964, Terrill1982}.  Whereas the above works assume static channels, other investigations have examined channels and cylinders with deforming walls as a function of time\cite{Majdalani2002, Asghar2008,Asghar2010,MohyudDin2010,XinHui2011,Rashidia11,Azimi2014,Sushila2014}.  All of these studies rely on concept of weak permeability in which the flow velocity at the wall is considered to be proportional wall velocity and independent of the hydrostatic pressure.  
Despite the physical reality that flow across channels is typically driven by hydrostatic pressure, we know of no analytic work that examines walls in motion while transcending the assumption of weak permeability.  

In the present work we determine an exact solution for Stokes flow through a channel with permeable and uniformly moving walls.  To derive this solution, we utilize the memoryless nature of the Stokes equations.  A wall in motion at any given point in time will be equivalent to prescribing a velocity at the wall coupled with the hydrostatically driven velocity field for a fixed channel width.  We will solve this static problem and use it to construct stream lines for a wall in motion at any given point in time.  

With a channel flow solution, we may then calculate the flux between the channel and reservoir.  We make the assumption that fluid in the reservoir may also be exchanged through pathways other than the channel-reservoir barrier.  In the current model, we assume this second exchange of reservoir fluid happens a rate proportional to the amount of fluid in the reservoir.  Physically, such a model suggests either a leak in the reservoir leading outside of the considered system or some process that consumes the fluid such such as evaporation, combustion, transport, or metabolism.
To close the relationship between reservoir pressure, and the amount of material within the reservoir, we consider a model in which the reservoir pressure is proportional to the amount of fluid within the reservoir; we note that more complicated closures may be incorporated, however do not consider these in the present work. 

With these assumptions, we model the homogeneous reservoir pressure $p_{res}$ as 
\begin{equation}
\dot{p}_{res} = -\gamma p_{res} + \alpha j(t,p_{res}),\label{eqn:ogmod}
\end{equation}
where $\gamma$ is the rate of extraction or metabolism, $j$ is the material flux entering or leaving the reservoir across the channel, and $\alpha$ is a constant of proportionality relating the volume flux to the rate of change of pressure within the reservoir.  The parameter $\alpha$ may be thought of as a stiffness coefficient of the reservoir; one simple physical interpretation is the pressurization of a piston with a spring in an attempt to return the system to an equilibrium volume.  The material flux may be derived by solving a system of fluid flow equations within the channel and then calculating the fluid flux across the boundary.  We assume that the channel has a uniform width that changes in time.  Along with pressure differences between the reservoir and the inlet/outlet conditions on the channel, the change in channel width will produce flux across the channel-reservoir barrier.

Having found a solution to the fluid flow equations, we then determine the flux between the reservoir and channel at any point in time and numerically solve our model for the reservoir pressure.  We find that changes in the amplitude and frequency of pumping can have significant effects on the amount of extracted fluid; the effects are shown to depend upon the channel length, the properties of the reservoir, and the system temperature.  Furthermore, we determine that there is a unique value of $\alpha$ that optimizes material extraction; such a result may have impact both on understanding the biological systems mentioned above as well as on filtration applications.


\section{Problem statement and solution}
\begin{figure}
\includegraphics[clip=true, trim=2cm 1cm 2cm 1cm, width=9cm]{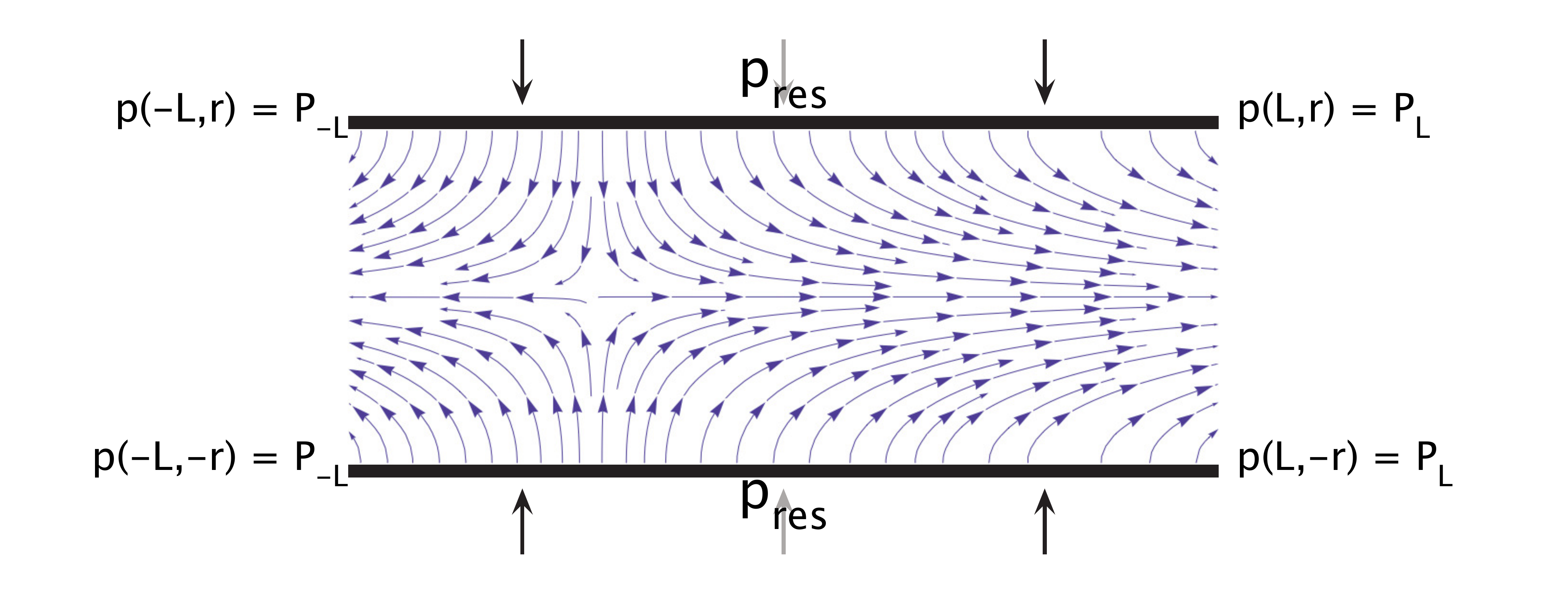}
\caption{A schematic of the physical set up.  The channel walls are moving toward each in the direction of the arrows and the pressure is fixed at the corner positions.  The pressure exterior to the channel is uniform.}
\label{fig:schematic}
\end{figure}

We begin this work by determining a formal power series solution for the flow described in the introduction; below we will utilize this solution in our model for $p_{res}$.  In two dimensions, the problem of Stokes flow with permeable boundary conditions and moving walls can be stated as
\begin{align}
&p(-L,\pm r(t)) = P_{-L}, \quad p(L,\pm r(t)) = P_{L}\label{eqn:primarysystem1}\\
&\nabla p = \mu\Delta \vec{u}, \quad \nabla \cdot \vec{u} = 0,\\
&u(x,\pm r(t)) = 0,\\
&v(x,\pm r(t)) = \pm  [\kappa(p(x,\pm r(t))-p_{res}) + \dot{r}(t)], \label{eqn:normalbdryog}
\end{align}
where $p$ is the pressure field; $u,v$ are the axial and cross sectional velocity fields, respectively; $P_{\pm L}$ are the pressure boundary conditions at $x=\pm L$, respectively with $y=\pm r(t)$; $r(t)$ is the half channel width as a function of time (uniform in space); and $p_{res}$ is the (uniform in space) pressure in the region of space exterior to the channel, considered to be some reservoir of fluid (see fig. \ref{fig:schematic}).  We let $x$ and $y$ denote the axial and cross sectional coordinates, respectively.   The choice of Darcy's law along with no slip boundary conditions at the channel walls has been previously justified in the case of isotropic boundaries with small permeability \cite{Tilton2012}, and can be further justified in biological applications where water moves across epithelial cell membranes through water channels in the normal, but not tangential, direction resulting in an anisotropic permeability for the membrane wall.  $p_{res}$ may also be set to zero without loss of generality.  We also note that we may change the boundary conditions from pressure to averaged flow conditions, for example
\begin{align}
\int_{-r(t)}^{r(t)} u(\pm L,y)dy = \bar{U}_{\pm L}.
\end{align}
Finally, we will assume that $r$ is a periodic function in time with period $T$.

To compute the solution we first note that the problem may be divided into two simplified problems via the linearity of the Stokes equations.  We separate the pressure-driven flow from the wall-driven flow by noting that the solution to equations \ref{eqn:primarysystem1}-\ref{eqn:normalbdryog} may be written as a superposition of solutions to the system given the pressure driven flow
\begin{align}
&p_p(-L,\pm r(t)) = P_{-L}, \quad p(L,\pm r(t)) = P_{L}\label{eqn:primarysystem1a}\\
&\nabla p_p = \mu\Delta \vec{u}_p, \quad \nabla \cdot \vec{u}_p = 0,\\
&u_p(x,\pm r(t)) = 0,\\
&v_p(x,\pm r(t)) = \pm  [\kappa(p_p(x,\pm r(t))-p_{res})], \label{eqn:normalbdryoga}
\end{align}
and the wall driven flow 
\begin{align}
&p_w(-L,\pm r(t)) = 0, \quad p_w(L,\pm r(t)) = 0\label{eqn:primarysystem1b}\\
&\nabla p_w = \mu\Delta \vec{u}_w, \quad \nabla \cdot \vec{u}_w = 0,\\
&u_w(x,\pm r(t)) = 0,\\
&v_w(x,\pm r(t)) = \pm  [\kappa(p_w(x,\pm r(t))) + \dot{r}(t)], \label{eqn:normalbdryogb}
\end{align}
where 
\begin{align}
\left( \begin{array}{c}
p \\
u \\
v \end{array} \right) = 
\left( \begin{array}{c}
p_p \\
u_p \\
v_p \end{array} \right)+\left( \begin{array}{c}
p_w \\
u_w\\
v_w \end{array} \right).
\end{align}
The pressure driven flow (equations \ref{eqn:primarysystem1a}-\ref{eqn:normalbdryoga}) has been well studied; an exact solution \cite{Herschlag:15} and many approximations for small permeability at non-zero Reynolds number can be found in the literature (see for example refs. \onlinecite{Haldenwang2007,Tilton2012}).

We will therefore restrict our focus to equations \ref{eqn:primarysystem1b}-\ref{eqn:normalbdryogb}.  We begin by nondimensionalizing this system at a fixed time with nondimensional variables
\begin{align}
\T{x} = x/r(t), \quad \T{y} = y/r(t),\\
\T{u} = u/\dot{r}(t), \quad \T{v} = v/\dot{r}(t)\\
\T{p} = p\kappa/\dot{r}(t).
\end{align}

Dropping the tildes and the subscript specifying wall driven flow, we arrive at the nondimensional equations
\begin{align}
&p(-\ell,\pm 1) = 0, \quad p(\ell,\pm 1) = 0,\label{eqn:nondimdop1}\\
&\nabla p = \mathcal{A}\Delta \vec{u}, \quad \nabla \cdot \vec{u} = 0,\\
&u(x,\pm 1) = 0,\\
&v(x,\pm 1) = \pm(p(x,\pm 1)+1),\label{eqn:nondimdop4}
\end{align}
where
\begin{align}
\ell = \frac{L}{r(t)},\quad 
\mathcal{A} = \frac{\mu \kappa}{r(t)}.
\end{align}
As previously noted, the memoryless nature of Stokes flows allows us to simply determine a solution for any fixed value of time.

\subsection{Solving the nondimensionalized system}
\label{sec:recursivesol}
To solve equations \ref{eqn:nondimdop1}-\ref{eqn:nondimdop4}, we will develop a formal expansion, letting
\begin{align}
\left( \begin{array}{c}
p \\
u \\
v \end{array} \right) = 
\sum_{n=1}^\infty \left( \begin{array}{c}
p_n \\
u_n \\
v_n \end{array} \right),
\end{align}
in which the normal velocity condition for the $n$th term satisfies the Darcy relationship for the residual pressure of the $(n-1)$th term.  To start the recursive procedure, we let the $n=1$ term satisfy the boundary condition given by the wall motion.  In other words, the boundary conditions for the normal velocity in each term are
\begin{align}
v_n(x,\pm 1)=\begin{cases} 
\pm 1, & n=1\\
\pm p_{n-1}(x,\pm 1), & n>1,
\end{cases}
\end{align}
which are used as boundary conditions for the set of equations given by the system
\begin{align}
&p_n(-\ell,\pm 1) =0, \quad p_n(\ell,\pm 1) = 0,\label{eqn:beginitsys}\\
&\nabla p_n = \mathcal{A}\Delta \vec{u_n}, \quad \nabla \cdot \vec{u_n} = 0,\\
&u_n(x,\pm 1) = 0,\label{eqn:enditsys}
\end{align}
for all $n\in\mathbb{N}$.

Below we will demonstrate that this recursive formula leads to a formal power series solution for $(p,u,v)$.  We will then investigate the effects of the parameters $\mathcal{A}$ and $\ell$ have on the convergence properties of this expansion.

\subsubsection{Inductive procedure to determine the formal series expansion}
The first order term (at $n=1$) is equivalent to the zeroth order expansion of the Berman problem \cite{Berman1953}.  It has the solution
\begin{align}
u_1(x,y) &= \frac{3 x}{2}\left(y^2-1\right),\label{eqn:u1}\\
v_1(x,y) &= \frac{1}{2}(3y-y^3),\\
p_1(x,y) &= \frac{3\mathcal{A}}{2}(x^2-y^2-\ell^2+1).\label{eqn:p1}
\end{align}
The normal velocity boundary condition for the second element of the expansion is then
\begin{align}
v_2(x,\pm 1) &= \frac{3\mathcal{A}}{2}(x^2-\ell^2).
\label{eqn:v2}
\end{align}
We note that we have a constant plus a quadratic term in $x$.  Due to the linearity of the Stokes equations, we can separate the two terms on the right hand side of equation \ref{eqn:v2}, and note that the constant boundary velocity term, $3\mathcal{A} \ell^2/2$, already gives rise to a solution promotional to the Berman solution presented in equations \ref{eqn:u1}-\ref{eqn:p1}.  As shown below, the $O(x^2)$ term found in equation \ref{eqn:v2} will result in a residual pressure that is a fourth order polynomial in $x$ with exclusively even coefficients.  More generally, we will show that an $n$th degree polynomial of $x^{2}$ for the $v_n$ velocity at the boundary will lead to a residual pressure that is an $(n+1)$th degree polynomial of $x^2$.  We will show this by induction and will also present a simple algorithm for constructing the higher order systems.

\begin{proposition}
If equations \ref{eqn:beginitsys}-\ref{eqn:enditsys} are closed with a normal wall velocity prescribed by an $n$th-degree polynomial of $x^2$,
$$v_n(x,\pm 1) = \pm \sum_{j=0}^{n} a_j x^{2 j},$$
then the pressure $p_n$ at the boundary will be an $(n+1)$th degree polynomial of $x^2$,
$$p_n(x,\pm 1) = \pm \sum_{j=0}^{n+1} b_j x^{2 j}.$$
\label{prop:bcrec}
\end{proposition}

To prove the proposition by induction, we first note that the base case has already been satisfied (equations \ref{eqn:u1}-\ref{eqn:p1}).  Supposing the inductive hypothesis holds true, we may note that each individual term in the polynomial for $v_n$ with $j<n-1$ leads to a residual pressure that is a polynomial of degree at most $(n-1)$ of $x^2$ at the boundary.  Therefore we must demonstrate that 
\begin{align}v_n(x,\pm 1) = \pm a_n x^{2 (n-1)}\label{eqn:proofsimpbdry}\end{align}
leads to a residual pressure that is a polynomial of $n$ of $x^2$ at the boundary.  We set $a_n=1$ without loss of generality, due to the linearity of the equations.

Let $\psi_n$ be the flow potential such that $u_n=\partial_y\psi_n$ and $v_n=-\partial_x\psi_n$.  Due to the symmetry of the boundary conditions on the pressure and velocity at the boundaries, we assume that the flow profile $u_n$ is odd in $x$ and even in $y$, that $v_n$ is even in $x$ and odd in $y$, and that $p_n$ is even in both $x$ and $y$.  Next, let us suppose there exists a finite power series such that 
\begin{align}
\psi_n = \sum_{j=0}^{(n+1)}\sum_{i=0}^{(n+1-j)} c_{ij} x^{2 i+1} y^{2 j+1}.
\label{eqn:ansatz}
\end{align}
This ansatz is natural in that it (i) ensures the symmetry conditions listed above, (ii) may satisfy the boundary conditions.  We note that should this ansatz give rise to a solution, then the pressure will be given by the indefinite integral
\begin{align}  
p(x,y) &= \int \Delta\partial_y \phi_n dx + f(y),
\end{align}
with an unknown constant of integration $f$ that is a function of $y$ alone.   By directly integrating equation \ref{eqn:ansatz} and then setting $y=1$, one obtains that the pressure at the boundary is a polynomial in $x$ of degree $2(n+1)$.  The symmetric boundary condition at $y=-1$ is satisfied due to the symmetry of the ansatz presented in equation \ref{eqn:ansatz}.

We then check condition (ii), which is equivalent to showing that the ansatz leads to a solution to the system given in equations \ref{eqn:beginitsys}-\ref{eqn:enditsys}, or that 
\begin{align}
\Delta \Delta \psi_n &= 0,\\ 
-\partial_x \psi_n(x,\pm 1) &= \pm x^{2(n-1)},\\
\partial_y \psi_n(x,\pm 1) &= 0.
\end{align}
The only unknowns left in the system are the $c_{ij}$'s, of which there are $(n+3)(n+2)/2$.
Plugging $\psi_n$ into the biharmonic equation and equating powers $x$ and $y$ results in $(n+1)n/2$ equations described by 
\begin{align}
P_4^{(2i+3)} c_{(i+1)(j-1)}&+2{P_2^{(2i+1)}} {P_2^{2j+1}} c_{ij}\nonumber\\
&+P_4^{(2j+3)} c_{(i-1)(j+1)}=0,\label{eqn:sys1}
\end{align}
for $j\in\{1,n\}$ and $i\in\{1,n+1-j\}$ due to the fact that when $i=0$ or $j=0$, the linear terms in $x$ and $y$ (respectively) vanish upon the first application of the Laplacian.  We also note that $P_k^n$ represents the $k$-permutations of $n$.  The normal velocity condition and no slip condition require that 
\begin{align}
-&\sum_{j=0}^{n+1-i} (2i+1) c_{ij} = \delta_{in},\label{eqn:sys2}\\
&\sum_{j=0}^{n+1-i} (2j+1) c_{ij} = 0.\label{eqn:sys3}
\end{align}
for $i\in\{0,n+1\}$ and where $\delta_{ij}$ denotes the Kronecker-$\delta$ function.  This system leads to an additional $2(n+2)$ equations.  According to the above calculations there appears to be one more equation than unknown, hinting that the system may be over determined. However, there is a trivial redundancy when $i=n+1$ in equations \ref{eqn:sys2} and \ref{eqn:sys3} which give, respectively,
\begin{align}
(2n+3)c_{(n+1)0} &= 0,\text{ and}\\
c_{(n+1)0} &= 0.
\end{align}

Thus we have reduced the system to a square matrix equation.  Next we will show the nonsingularity of the matrix by developing a recursion formula to directly solve the system. To do this we note from above that $c_{(n+1)0}=0$ and that we have $(n+2-i)$ unknowns with two equations for every fixed value of $i$ using equations \ref{eqn:sys2} and \ref{eqn:sys3}.  We have already determined the case where $i=n+1$.  For $i=n$ we obtain the $2\times 2$ dimensional system
\begin{align}
\left(\begin{array}{cc} 1 & 3 \\ (2n+1) & (2n+1) \end{array} \right)\left(\begin{array}{c} c_{n0} \\ c_{n1}\end{array} \right)= \left(\begin{array}{c} 0 \\ -1\end{array} \right),
\end{align}
which has solution
\begin{align}
c_{n0} = -\frac{3}{2 (2n+1)}, \quad c_{n1} = \frac{1}{2 (2n+1)}.\label{eqn:hot}
\end{align}
Proceeding recursively, we assume that we have found $c_{ij}$ for all values of $i'$ greater than some fixed $i$ and $j=\{1,...,n+1-i'\}$ for each value of $i'>i$.  We seek to demonstrate that equation \ref{eqn:sys1} can be used to determine $c_{ij}$ for $j\geq 2$, thereby reducing the $(n+2-i)$ to two parameters (namely $c_{i0}$ and $c_{i1}$).  That is equations \ref{eqn:sys2} and \ref{eqn:sys3} are reduced to
\begin{align}
\left(\begin{array}{cc} 1 & 3 \\ (2i+1) & (2i+1) \end{array} \right)\left(\begin{array}{c} c_{i0} \\ c_{i1}\end{array} \right) + \left(\begin{array}{c} C_1 \\ C_2 \end{array} \right)= \left(\begin{array}{c} 0 \\ 0\end{array} \right),
\end{align}
with
\begin{align}
C_1 = \sum_{j=2}^{n+1-i} (2j+1) c_{ij},\quad
C_2 = \sum_{j=2}^{n+1-i} (2i+1) c_{ij}.
\end{align}
This system is linearly independent.  Thus the entire system is invertible and a unique polynomial exists.  The remaining claim is straightforward as we have equation \ref{eqn:sys1} for all values of $i\geq1$, and thus we can find $c_{ij}$ (for $1<j<n+1-i$) by setting it as a function of $c_{(i+1)(j-1)}$ and $c_{(i+2)(j-2)}$
\begin{align}
c_{ij}=\frac{P_4^{(2i+3)}c_{(i+2)(j-2)}+2{P_2^{(2i+3)}} {P_2^{2j-1}} c_{(i+1)(j-1)}}{P_4^{(2j+1)}}.
\end{align}
This procedure may then be repeated up to the case in which $i=1$ and we have thus found a recursive algorithm for the solution to equations \ref{eqn:beginitsys}-\ref{eqn:enditsys} with normal velocity boundary condition given by equation \ref{eqn:proofsimpbdry}.  This completes the proof of proposition \ref{prop:bcrec}. $\blacksquare$

Because the pressure $p_n$ is the boundary condition for the normal velocity $v_{n+1}$, this proposition allows us to recursively construct a formal power series solution for equations  \ref{eqn:nondimdop1}-\ref{eqn:nondimdop4}, provided that the expansion converges.  In the next subsection we will examine the convergence properties of this expansion conditioned on the nondimensional parameters $\mathcal{A}$ and $\ell$.

\subsubsection{Convergence}
To demonstrate the convergence of the series expansion, we seek an upper bound  on the radius of convergence.
%

  First we note that $\psi_n$ may be written in terms of the highest order term in $x$, given by equation \ref{eqn:hot} as
\begin{align}
\psi_n(x,y) = \mathcal{A}^n\left(\frac{y^3 x^{2n+1}-3 y x^{2n+1}}{2(2n+1)}+o(x^{2n-1}))\right).\label{eqn:psin}
\end{align}
Letting the potential function be $\Psi = \sum_{n=1}^\infty \psi_n$, we use the ratio test to find an upper bound on the radius of convergence in $x$, and find that $|x|<\mathcal{A}^{-1/2}$, which provides an upper bound for $\ell$ based on $\mathcal{A}$.  

\begin{figure}

\includegraphics[width=9cm]{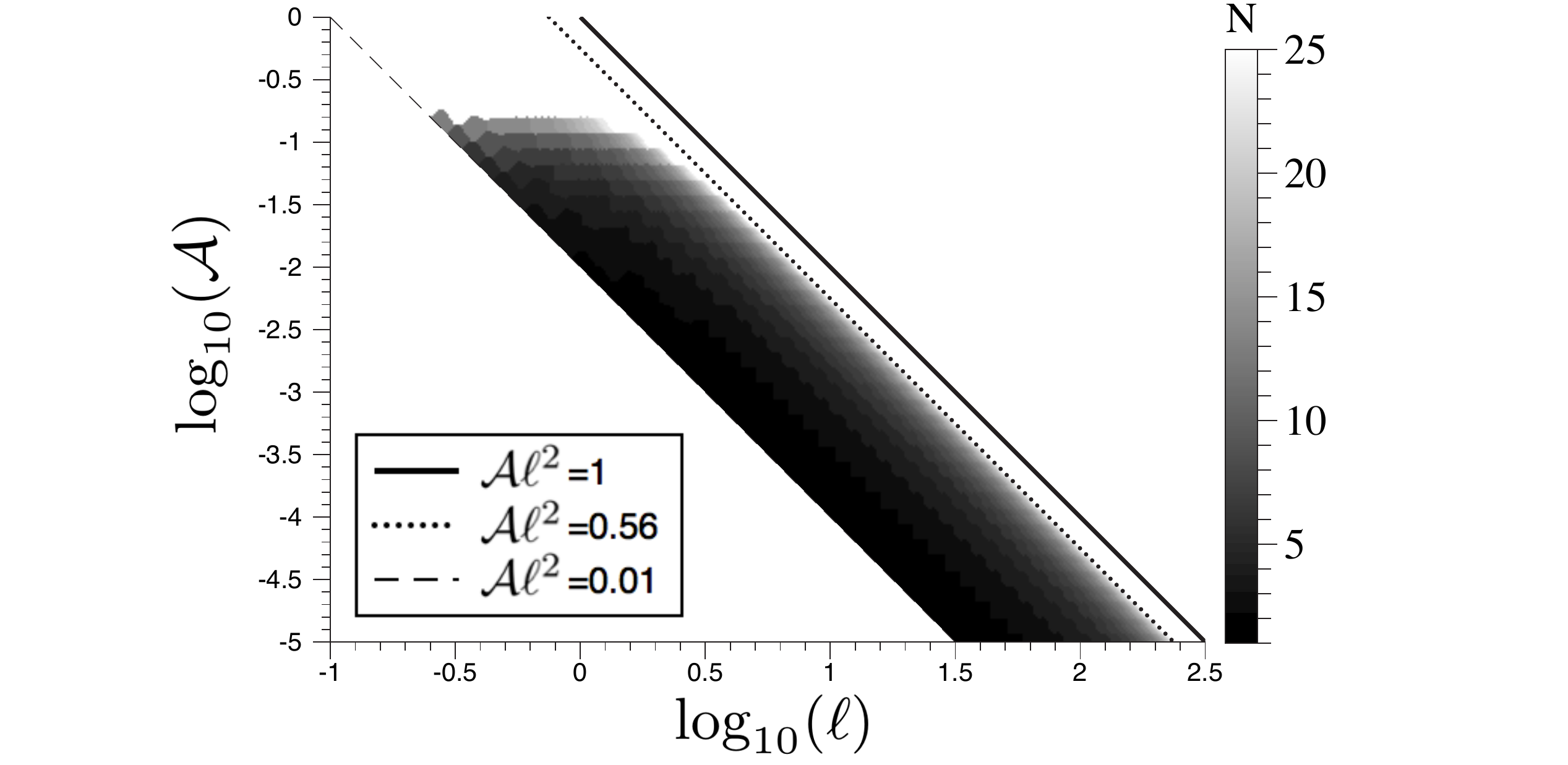}\\(a)
\includegraphics[width=9cm]{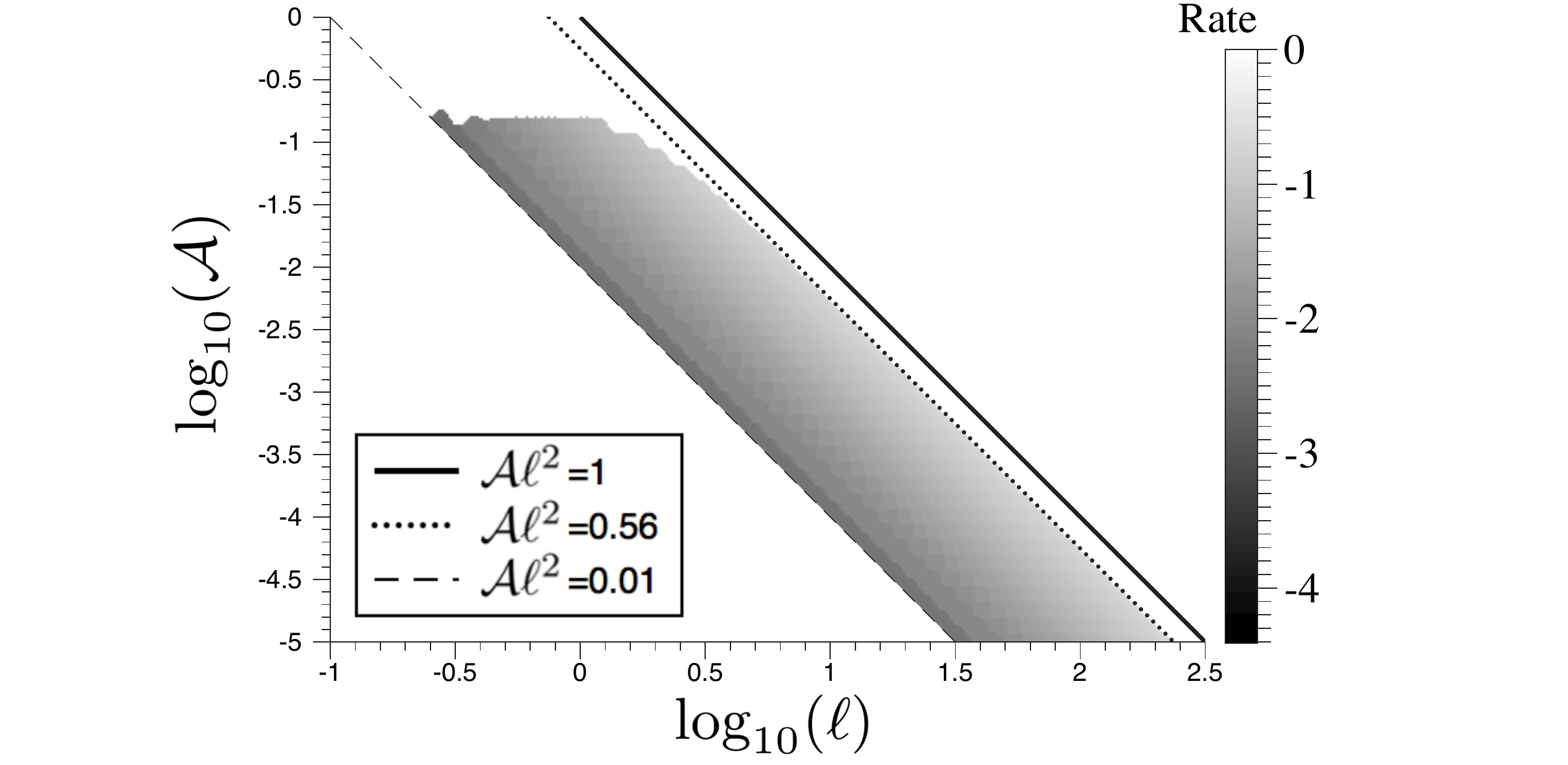}\\(b)
\caption{The number of recursive steps needed to obtain a relative error of $10^{-4}$ is displayed (a).  The maximum number of steps taken is limited to 25.  We have only investigated parameters above the dashed line which displays $\mathcal{A}\ell^2 = 0.01$.  In order to determine the rate of convergence, and the existence of other parameters for which the recursive solution converges, we plot the rate of convergence given by a linear fit of $\log_{10}(\epsilon(N))$ as a function of $N$. Regions colored white (and above the dashed line) signify diverging power series (b).  We plot the line $\mathcal{A}\ell^2 = 1$ which is an analytically derived upper bound of this product and $\mathcal{A}\ell^2 = 0.56$ which is the numerically derived upper bound.}
\label{fig:convg}
\end{figure}

Next, we analyze the convergence numerically by examining a variety of values for $\mathcal{A}$ and $\ell$. Due to the criteria of the ratio test that $\mathcal{A}\ell^2<1$, we plot the convergence regimes parametrized by $\log(\mathcal{A})$ and $\mathcal{A}\ell^2$.  There will be no error between the approximation $\Psi_N =\sum_{i=1}^N\psi_n$ and the exact solution for all equations except the normal velocity boundary condition, presented in equation \ref{eqn:nondimdop4}.  Thus we define the error to be $||V_{N}-(1+P_{N})||_\infty$, where $V_N=\sum_{n=1}^N v_n$ and  $P_N=\sum_{n=1}^N p_n$.  We define the relative error to be the error normalized by sup norm of the average  between $P+1$ and $V$ given as 
\begin{align}
\epsilon(N)\equiv \frac{||V_{N}-(1+P_{N})||_\infty}{\frac{1}{2}(||V_{N}||_\infty+||(1+P_{N})||_\infty)}.
\end{align}
We vary $\mathcal{A}$ between $10^{-5}$ and 1 on a log scale and $\mathcal{A}\ell^2$ between $0.01$ and $1$.  We increase $N$ until either the relative error is below a threshold (set to be $10^{-4}$) or until $N=25$.  We record the regions of convergence along with the depth of recursion needed to find convergence in figure \ref{fig:convg}(a), and display the rate of convergence in figure \ref{fig:convg}(b).  We note that for figure \ref{fig:convg}(b), we do not consider whether or not the error threshold has been achieved but instead report the order of the convergence. 

We find that the upper bound of $\mathcal{A}\ell^2<1$ is not sharp over all parameter regimes.  Indeed, the error is increasing between truncation value $N=24$ and $N=25$ for all values of $\mathcal{A}\ell^2>0.56$.  In the current parameter runs we have found an approximate upper bound on $\mathcal{A}$ of $10^{-5/8}$.  We did not numerically investigate a sharper restriction as permeabilities greater than $\mathcal{A}=10^{-5/8}$ require an exceedingly short channel, with channel length less than one fifth of the channel width, in order for the recursive solution to converge.  Thus the resulting formal solution is unlikely to be of any aid in modeling physical systems in this parameter regime.  Our numerical results also reveal that the recursive technique will not converge when (equivalently) the length is too long or the channel width too narrow by bounding $\ell$ for fixed $\mathcal{A}$.

%

\section{Solution properties}
Having determined a formal series solution and numerically investigated convergence properties, we now investigate the properties of our solution in the context of previous work (see below).  We begin by analyzing the assumption of weak permeability.  We continue by determining the proximity of our solution to a self-similar solution in the axial direction, as much of the work in this field has investigated solutions with this property.

\subsection{A comparison with the assumption of weak permeability}
As previously noted, there has been a considerable amount of work examining the assumption of weak permeability (see, for example, refs. \onlinecite{Majdalani2002, Asghar2008,Asghar2010,MohyudDin2010,XinHui2011,Rashidia11,Azimi2014,Sushila2014}).  Weak permeability assumes that the velocity at the wall is proportional, but not equal, to the wall velocity at all points along the tube.  That is, the normal boundary velocity (eqn. \ref{eqn:normalbdryog}) is given by
\begin{align} v(x,\pm r(t)) = \dot{r}(t)/c,\end{align}
where the permeance coefficient $c$ represents the ratio between the wall speed and the flow velocity at the wall.  In nondimensional units this boundary condition becomes
\begin{align} v(x, \pm 1) = 1/c. \end{align}

\begin{figure}
\includegraphics[width=9cm]{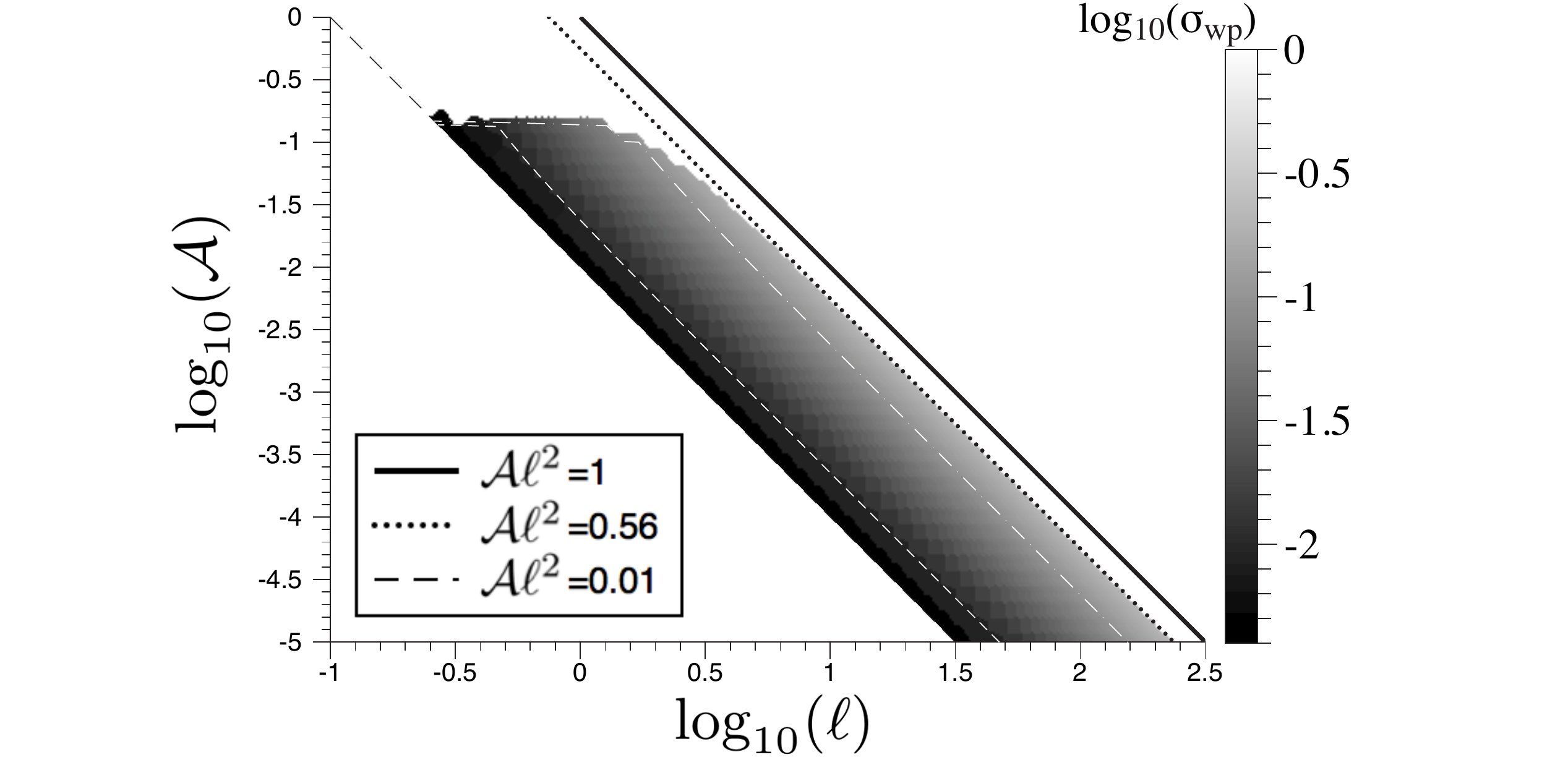}
\caption{The variation $\sigma_{wp}$ away from the weakly permeable limit.  Note that for all values of the nondimensional permeability, $\mathcal{A}$, the assumption of weak permeability breaks down as the ratio between channel length and width, $\ell$ is sufficiently large.  Additionally, we display the level set curve on which the variation is 10\% (white dot-dashed line) and 20\% (white dashed line).}
\label{fig:weakp}
\end{figure}

\begin{figure}
\includegraphics[width=7cm]{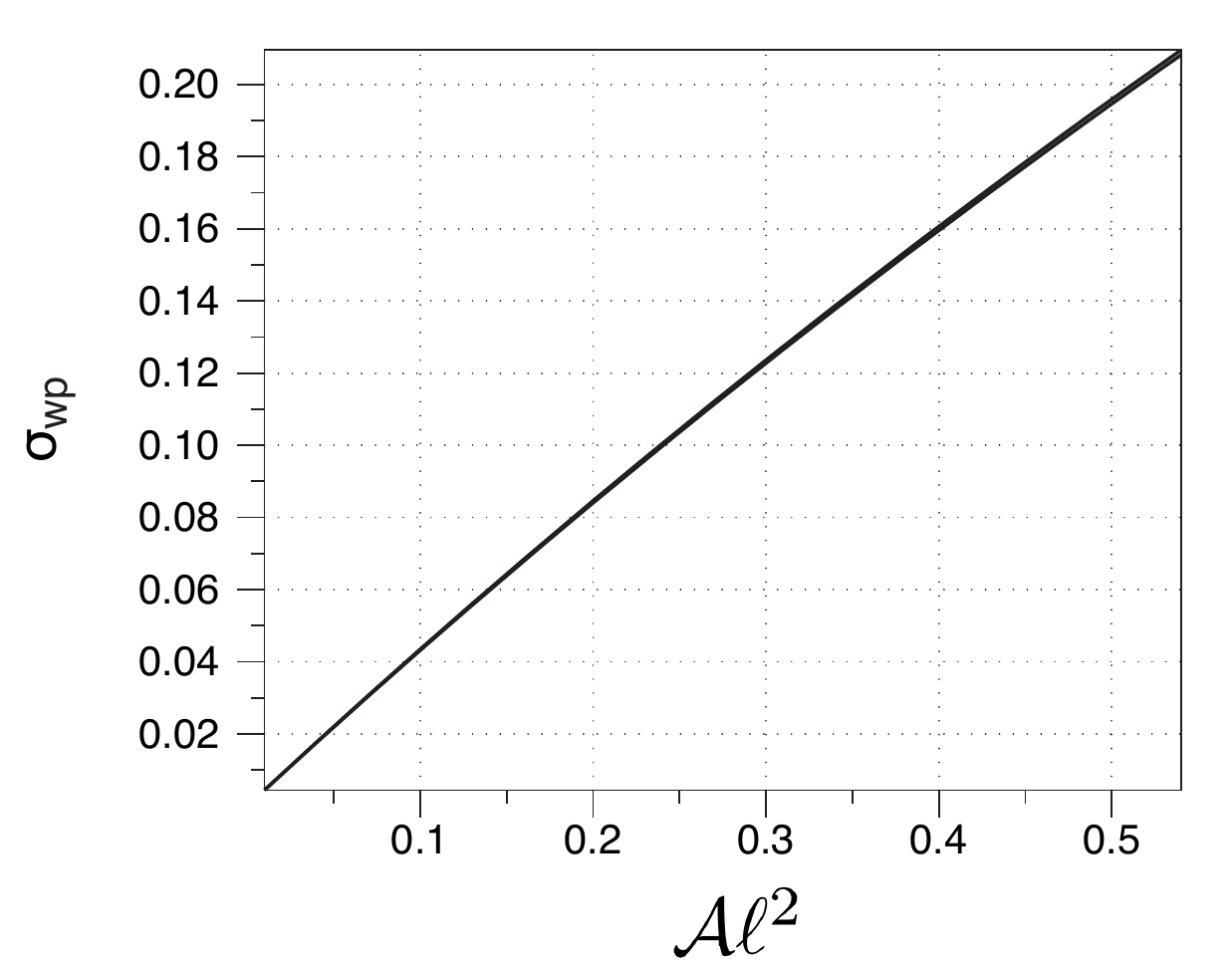}
\caption{Variation $\sigma_{wp}$ as a function of $\mathcal{A}\ell^2$ is in roughly invariant for $\mathcal{A}\in[10^{-6},3.16\times10^{-3}]$.  The profile is nearly invariant for these small values of $\mathcal{A}$.}
\label{fig:invarsigb}
\end{figure}

It is natural to ask how our model, driven by a hydrostatic pressure difference across the channel walls, compares to a model assuming weak permeability.  To analyze this, we ignore the pressure driven flow, use our recursive solution to equations \ref{eqn:nondimdop1}-\ref{eqn:nondimdop4} and approximate 
\begin{align}\frac{1}{c} = \bar{v}\equiv \frac{1}{2\ell}\int_{-\ell}^{\ell} v(x,1)dx. \end{align}
We then determine the relative variation away from the assumption of weak permeability as 
\begin{align}
\sigma_{wp} \equiv \frac{1}{|\bar{v}|}\sqrt{\frac{1}{2 \ell}\int_{-\ell}^{\ell} (v(x,1)-\bar{v})^2 dx}.
\end{align}
$\sigma_{wp}$ is constructed so that it is invariant under linear spatial transformations.  As shown in figure \ref{fig:weakp}, the assumption of weak permeability breaks down for all values of the nondimensional permeability, $\mathcal{A}$, as the ratio between channel length and width, $\ell$, increases.  Additionally, for values of $\mathcal{A}<0.1$, the variation $\sigma$ is roughly fixed when a value for $\mathcal{A}\ell^2$ is fixed.  To see this concretely we overlay plots of $\sigma_{wp}$ as a function of $\mathcal{A}\ell^2$ for values of $\mathcal{A}\in[10^{-6},3.16\times10^{-3}]$ in figure \ref{fig:invarsigb}.

\subsection{Flow profile variation along the axial direction}
Perhaps the most common technique in developing analytic theory of flow through permeable channels and pipes is to assume a self similar profile along the axial direction.  The present series expansion is not self similar (see equation \ref{eqn:psin}).  Nevertheless, one may ask how the profiles of the field variables  change with respect to the axial position.  

\begin{figure}
\includegraphics[trim = 0mm 0mm 40mm 0mm, clip, width=8cm]{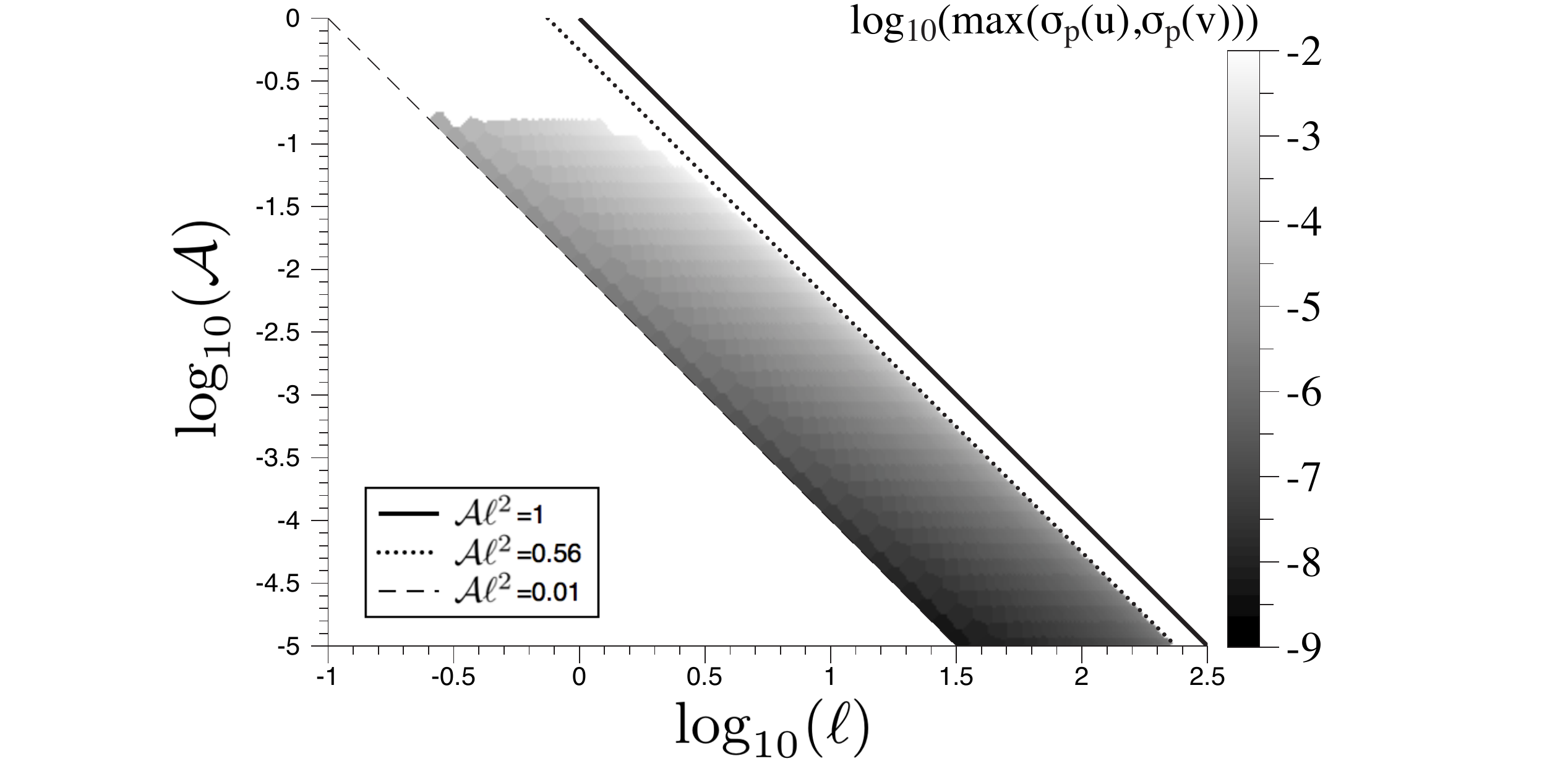}(a)\\
\includegraphics[trim = 0mm 0mm 40mm 0mm, clip, width=8cm]{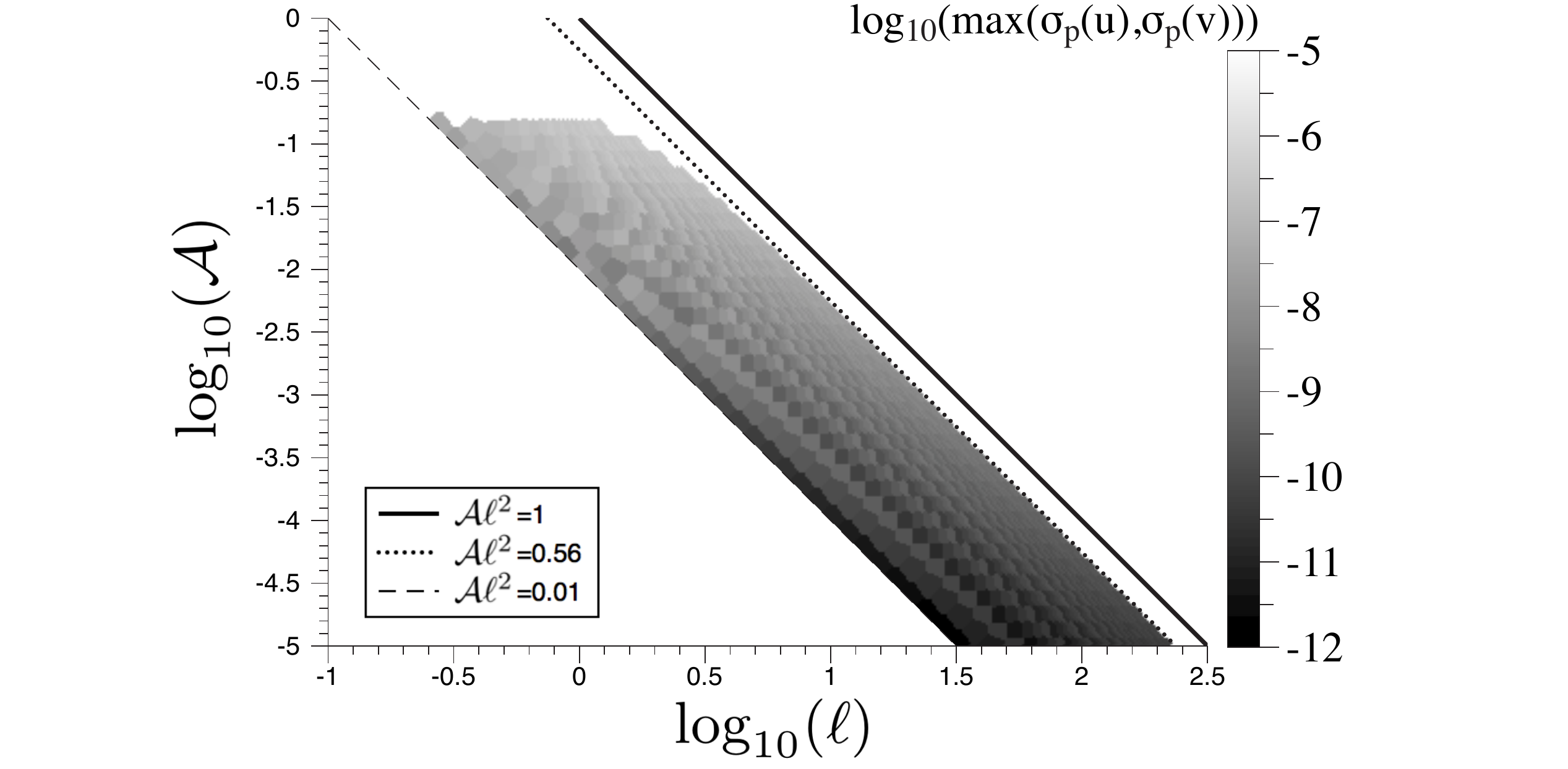}(b)
\caption{We plot $\max(\sigma_{\text{p}}(u),\sigma_{\text{p}}(v))$ for the asymptotic expansion presented in equations \ref{eqn:asymu} and \ref{eqn:asymv} (a).  We also plot $\max(\sigma_{\text{p}}(u),\sigma_{\text{p}}(v))$ for the solutions that have converged numerically with an error tolerance of $10^{-4}$ (b).}
\label{fig:siguv}
\end{figure}

We proceed both analytically and numerically.  For the analytic aspect, we first take a different nondimensionalization of equations \ref{eqn:primarysystem1}-\ref{eqn:normalbdryog} by letting
\begin{align}
\T{x} = x/r(t), \quad \T{y} = y/r(t),\\
\T{u} = u/\dot{r}(t), \quad \T{v} = v/\dot{r}(t)\\
\T{p} = p  \kappa/\mathcal{A}\dot{r}(t).
\end{align}
Setting the boundary pressures to be zero, we arrive at a new system of nondimensional equations (dropping tildes)
\begin{align}
&p(-\ell,\pm 1) =0, \quad p(\ell,\pm 1) = 0,\label{eqn:req1new}\\
&\nabla p = \Delta \vec{u}, \quad \nabla \cdot \vec{u} = 0,\\
&u(x,\pm 1) = 0,\\
&v(x,\pm 1) = \pm (\mathcal{A} p(x,\pm 1)+1).\label{eqn:req3new}
\end{align}
This system may be solved using an algorithm identical to that presented in section \ref{sec:recursivesol}, with the exception of an extra-power of $\mathcal{A}$ in the expansion for $p$ which is accounted for in the new nondimensional units.  Thus we may now view the formal series expansion as an asymptotic expansion about small values for $\mathcal{A}$.  This is a reasonable restriction given that the series only converges for sufficiently small values of $\mathcal{A}$.  We then note that we may approximate our solution with the first two terms in the asymptotic expansion, where the zeroth order term, $O(\mathcal{A}^0)$, corresponds to Berman's solution.  This expansion predicts that
\begin{align}
u(x,y) \approx& \frac{3}{20} x \left(-1+y^2\right) \times\nonumber \\
&\left(10+\mathcal{A} \left(1-15 \ell^2+5 x^2-5 y^2\right)\right)\label{eqn:asymu},\\
v(x,y) \approx& \frac{1}{20} y \Big[-10 \left(-3+y^2\right)+\nonumber\\
&3 \mathcal{A} \Big\{5 \ell^2 \left(-3+y^2\right)-\nonumber\\
&5 x^2 \left(-3+y^2\right)+\left(-1+y^2\right)^2\Big\}\Big]\label{eqn:asymv}.
\end{align}
Although these approximations are, strictly speaking, not self similar profiles, the deviation is small.  This can be seen by considering the variance in the profile
\begin{align}
\sigma_{\text{p}}(w)\equiv \sup_{x_1,x_2\in[0,\ell]} \inf_{\beta\in\mathbb{R}} \sqrt{\frac{\int_{-1}^{1} (w(x_1,y)-\beta w(x_2,y))^2 dy}{\int_{-1}^{1} w(x_1,y)^2 dy}}.
\end{align}
Because the integrands are polynomials, we can analytically determine $\sigma_{\text{p}}(u)$ and $\sigma_{\text{p}}(v)$ for any given set of parameters $\mathcal{A}$ and $\ell$.  In figure \ref{fig:siguv}, we plot $\max(\sigma_{\text{p}}(u),\sigma_{\text{p}}(v))$ and see that with the leading order asymptotic theory, the profile variation stays below  $5\times10^{-3}$.  We repeat the analysis numerically for profiles that have converged below the error tolerance.  To save on the numerical expense of directly calculating $\sigma_{\text{p}}(w)$, we approximate it by setting $\beta = \max_y(|w(x_1,y)/w(x_2,y)|)$ and consider the limit as $x_1\rightarrow\ell$ and $x_2\rightarrow0$.  Our choice for $\beta$ ensures that computed value will be an upper bound for $\sigma_p$; our choice for $x_1$ and $x_2$ is made because numerical test cases suggest this will yield the greatest difference in the profile shape.  We find that the profiles vary less as more terms in the series expansion are added.  All profile variations for the numerical results were bounded below $5\times10^{-7}$, which is a surprising result considering that the error tolerance on the numerical scheme was set to $10^{-4}$.  This may explain the fluctuations in figure \ref{fig:siguv}(b).  There is nothing \emph{a priori} to suggest the solution should form a self similar profile; indeed $\sigma_{\text{p}}(p)$ the pressure profile is not self similar in the pressure profile.  Nevertheless, we have shown that the profile may be well approximated by a self similar profile.

%

\subsection{Range of validity}
We now consider the types of systems for which the above solution may be valid.  We begin by remarking that as the channel becomes thinner, $\ell$ becomes larger for fixed length, and thus our solution will only be valid so long as $\inf_{t\in\mathbb{R}} r(t) \geq \sqrt[3]{0.56/\kappa \mu L}$.  For fixed $\kappa$ and $\mu$ this sets an upper bound on the allowable length scales of the channel.  In ref. \onlinecite{Tilton2012}, the authors perform a brief literature review and find that $\mathcal{A}$ is typically in the range of $10^{-6}$ and $10^{-14}$ nondimensional permeability.  This means that the maximal channel to tube length ratio ($\max_t \ell=\max_t L/r(t)$) must be somewhere in the range of $7.5\times10^2$ and $7.5\times10^6$ depending on the value of $\mathcal{A}$.  The solution will break down for fully compressed channels for which $r(t)\rightarrow 0$.

\section{An averaged flow model for material extraction}
Having determined a solution for a specified range of parameters above, we now return to the model for $p_{res}$ presented in equation \ref{eqn:ogmod}.  We adopt a periodic, symmetric-in-time pumping strategy such that
\begin{align}
r(t) = \bar{r} + A \sin(\omega t).
\end{align}
The flux in equation \ref{eqn:ogmod} may now be described as
\begin{equation}
j(t,p_{res}) = \int_{-L}^{L} \big(v(x,r(t); p_{res})-\dot{r}(t)\big) dx,\label{eqn:modelextract}
\end{equation}
where $v$ is the normal velocity from equations \ref{eqn:primarysystem1}-\ref{eqn:normalbdryog}.  We subtract the speed of the wall in the interior of the integral of equation \ref{eqn:modelextract} in order to obtain the flux across the wall, and then integrate to determine the total flow into the reservoir.  The parameter $\alpha$ relates the rate of reservoir volume change to the rate of pressure change.  One simple physical interpretation is the pressurization of a piston with a spring in an attempt to return the system to an equilibrium volume.  More complex models that relate volume flux to pressure are also possible, but we do not consider them in the present work.

We will denote the pressure driven flow (the solution of equations \ref{eqn:primarysystem1a}-\ref{eqn:normalbdryoga}) as $v_{press}(x)=v_p(x,r(t))$ and the wall motion-driven flow (the solution of equations \ref{eqn:primarysystem1b}-\ref{eqn:normalbdryogb}) as $v_{pump}(x)=v(x,r(t))$. From the solution to equations \ref{eqn:primarysystem1a}-\ref{eqn:normalbdryoga} \cite{Herschlag:15}, we determine that the normal velocity at $y=r(t)$ from the pressure driven flow is given by
\begin{align}
v_{press}(x,y=r(t)) = \kappa \frac{P_{L}+P_{-L}-2 p_{res}}{2}\frac{\cosh\left(\lambda(t) x/r(t) \right)}{\cosh(L\lambda(t)/r(t))}\nonumber \\
+\frac{P_{L}-P_{-L}}{2}\frac{\sinh\left(\frac{\lambda(t) x}{r(t)} \right)}{\sinh\left(L\lambda(t)/r(t)\right)},
\end{align}
where $\lambda(t)$ is the solution to the transcendental equation 
\begin{align}
\mathcal{A}(t)\equiv \frac{\kappa\mu}{r(t)} = \frac{1}{2}\left(\frac{1}{\cos^2(\lambda(t))}-\frac{\sin(\lambda(t))}{\lambda(t) \cos(\lambda(t))}\right).
\end{align}

To nondimensionalize equation \ref{eqn:modelextract}, we set
\begin{align}
 t = \tilde{t}/\gamma, \quad p_{res} = \alpha \bar{r}^2 \tilde{p}_{res}, \quad r(t) = \tilde{r}(t) \bar{r},
\end{align}
and obtain
\begin{align}
\dot{\tilde{p}}_{res} =& -\tilde{p}_{res}
 \frac{\tilde{r}(\tilde{t})}{\lambda(\tilde{t})}(\Xi- \Upsilon \tilde{p}_{res}(\tilde{t}))\tanh(\lambda(\tilde{t})\ell(\tilde{t}))
\nonumber\\
&+\Theta \Omega \cos(\Omega \tilde{t}) \int_{-\ell(\tilde{t})}^{\ell(\tilde{t})} \big(v_{\text{pump}}(x,1;\mathcal{A}(\tilde{t}),\ell(\tilde{t}))-1
\big) dx,\label{eqn:modelextractnondim}
\end{align}
where $v_{pump}$ is the (nondimensional) solution for equations \ref{eqn:nondimdop1}-\ref{eqn:nondimdop4} parametrized by $\mathcal{A}(\tilde{t})$ and $\ell(\tilde{t})$.  The nondimensional constants are defined as
\begin{align}
\Xi = \frac{\kappa (P_{L}+P_{-L})}{\gamma \bar{r}},\quad \Upsilon=\frac{2 \alpha \bar{r} \kappa}{\gamma},\quad \bar{\mathcal{A}} = \frac{\kappa \mu}{\bar{r}}, \\
\bar{\ell}= L/\bar{r},\quad
\Theta = \frac{A}{\bar{r}},\quad \Omega=\frac{\omega}{\gamma},
\end{align}
and the time varying nondimensional parameters are defined as
\begin{align}
\mathcal{A}(t) = \bar{\mathcal{A}}/\tilde{r}(\tilde{t}),\quad \ell(t) = \bar{\ell}/\tilde{r}(\tilde{t}).
\end{align}

We thus arrive at a linear first order differential equation characterized by six nondimensional parameters.  We may express the exact solution of this equation in terms of a convolution
\begin{align}
\tilde{p}_{res}(\tilde{t}) =&\tilde{p}_{res}(0) \exp\left(-\int_0^{\tilde{t}} g(s)\right)\nonumber\\
&+\exp\left(-\int_0^{\tilde{t}} g(s) \right)\ast f(\tilde{t}),\label{eqn:analyticsolpump}
\end{align}
where 
\begin{align}
g(\tilde{t}) \equiv& 1+\Upsilon \frac{\tilde{r}(\tilde{t})}{\lambda(\tilde{t})}\tanh(\lambda(\tilde{t})\ell(\tilde{t}))\\
f(\tilde{t}) \equiv& \Xi  \frac{\tilde{r}(\tilde{t})}{\lambda(\tilde{t})} \tanh(\lambda(\tilde{t})\ell(\tilde{t}))\\
&+\Theta \Omega \cos(\Omega \tilde{t}) \int_{-\ell(\tilde{t})}^{\ell(\tilde{t})} \big(v_{\text{pump}}(x,1;\mathcal{A}(\tilde{t}),\ell(\tilde{t}))-1
\big) dx.
\end{align}
By Floquet theory, a time-periodic solution exists, and since the coefficients are smooth, this solution is unique up to transient behavior.  Below we examine the effect of selected parameters on the mean value of the reservoir pressure and we show that pumping can indeed enhance the average material extraction. 

\subsection{Effect of varying $\Upsilon$ and optimal reservoir conditions}
We consider the effect of changes in $\Upsilon$, which is the ratio between the rate at which material pressurizes the system, and the rate at which the material decays in the reservoir.  A reservoir which does not change pressure with incoming fluid corresponds to $\Upsilon=0$ and an infinitely stiff reservoir corresponds to $\Upsilon\rightarrow\infty$.  For simplicity, we set $\Xi=0$, which means that there is no material in the reservoir in the absence of pumping.

We first consider the limit of small $\Upsilon$.  We note that the coefficients have period $2 \pi/\Omega$.  Suppose then that we have a periodic solution and expand the solution in the Fourier basis.  The zeroth mode gives the mean value of the solution which corresponds to the zeroth mode of $f(\tilde{t})$.  We note that average value of $f(\tilde{t})$ is the amount of fluid that enters and exits the reservoir after a contraction and expansion in the absence of metabolic activity, and therefore has mean zero due to the reversibility of the Stokes equations.  Thus periodic solutions for the reservoir pressure have zero mean.  This explains the case in which reservoir does not change its pressure with respect to volume.  

Next consider the limit of large $\Upsilon$. In this case, $f(\tilde{t})$ is small compared to $g(\tilde{t})-1$, i.e. $\Theta \Omega/\Upsilon=\epsilon\ll 1$ (since $\Xi=0$).  We have determined in the previous section that $f(\tilde{t})$ is absolutely bounded within its radius of convergence, and further note that $g(\tilde{t})>1$ for all time, which implies that the first term of the convolution is bounded by 1.  Letting $M$ be the absolute bound for $f(\tilde{t})$, we conclude that the convolution found in equation \ref{eqn:analyticsolpump} is absolutely bounded by $\epsilon M$.  Furthermore, the first term of equation \ref{eqn:analyticsolpump} decays to zero in long time.  It follows that the reservoir pressure goes to zero as $\epsilon\rightarrow0$ and $\tilde{t}\rightarrow\infty$.

We have left to inquire about the intermediate values of $\Upsilon$.  To numerically investigate the effect on $\Upsilon$, we consider the restriction in parameter space of bounding $-0.5<\Theta<0.5$.  As mentioned above, we consider realistic values for the nondimensional permeability, confining $\bar{\mathcal{A}}\in[10^{-14},10^{-6}]$.  These two constraints determine a maximal value for $\bar{\ell}$ of 265, to ensure we remain within the radius of convergence of the formal power series for $v_{\text{pump}}$.  With these constraints in mind, we proceed by setting $\Omega=10$, $\bar{\mathcal{A}}=10^{-6}$, $\bar{\ell}=200$, and vary $|\Theta|$ from 0.1 to 0.5, and $\Upsilon$ from $0$ to $0.05$.  We then determine the long time average of $\tilde{p}_{res}(\tilde{t})$,
\begin{align}
\langle\tilde{p}_{res}(\tilde{t})\rangle = \frac{\Omega}{2 \pi} \lim_{\tilde{t}\rightarrow\infty}\int_{\tilde{t}-\pi/\Omega}^{\tilde{t}+\pi/\Omega}\tilde{p}_{res}(\tilde{t}) dt.
\end{align}
We solve the system using Mathematica's built-in NDSolve routine \cite{mathematica}.  For the remainder of the paper, we truncate the formal series solution for $v_{pump}$ at $N=5$ since (i) the relative two norm is below 0.5\% and (ii) we have noticed no difference in the results presented below.  Results are shown in figure \ref{fig:presvup}.  For each value of $\Theta$, an optimal value of $\Upsilon$ for material extraction can be found between 0.016 and 0.021, which implies parameter regimes can be identified in which pumping generates maximal enhanced material extraction.  In other words, there is an $\Upsilon$ that maximizes material extraction.  

We continue by considering $\langle\tilde{p}_{res} \rangle$ as a function of $\Upsilon$ for $\Omega\in[10,100]$, $\bar{\ell}\in[100,200]$, and $\bar{\mathcal{A}}\in[10^{-8},10^{-6}]$ (see figure \ref{fig:presvup} (b)-(d), respectively); additionally, $\Theta=0.5$, $\Omega=10$, $\bar{\mathcal{A}}=10^{-6}$, and $\bar{\ell}=200$.  Our results indicate that the optimal $\Upsilon$ is insensitive with respect to $\Theta$ and $\bar{\mathcal{A}}$, changes with respect to $\bar{\ell}$, and is highly sensitive to changes in $\Omega$.  This suggests that one ought to design a reservoir with a particular frequency in mind and then vary $\Theta$ to change the amount of material extraction.

\begin{figure*}
\includegraphics[trim = 0mm 0mm 0mm 0mm, clip, width=7.5cm]{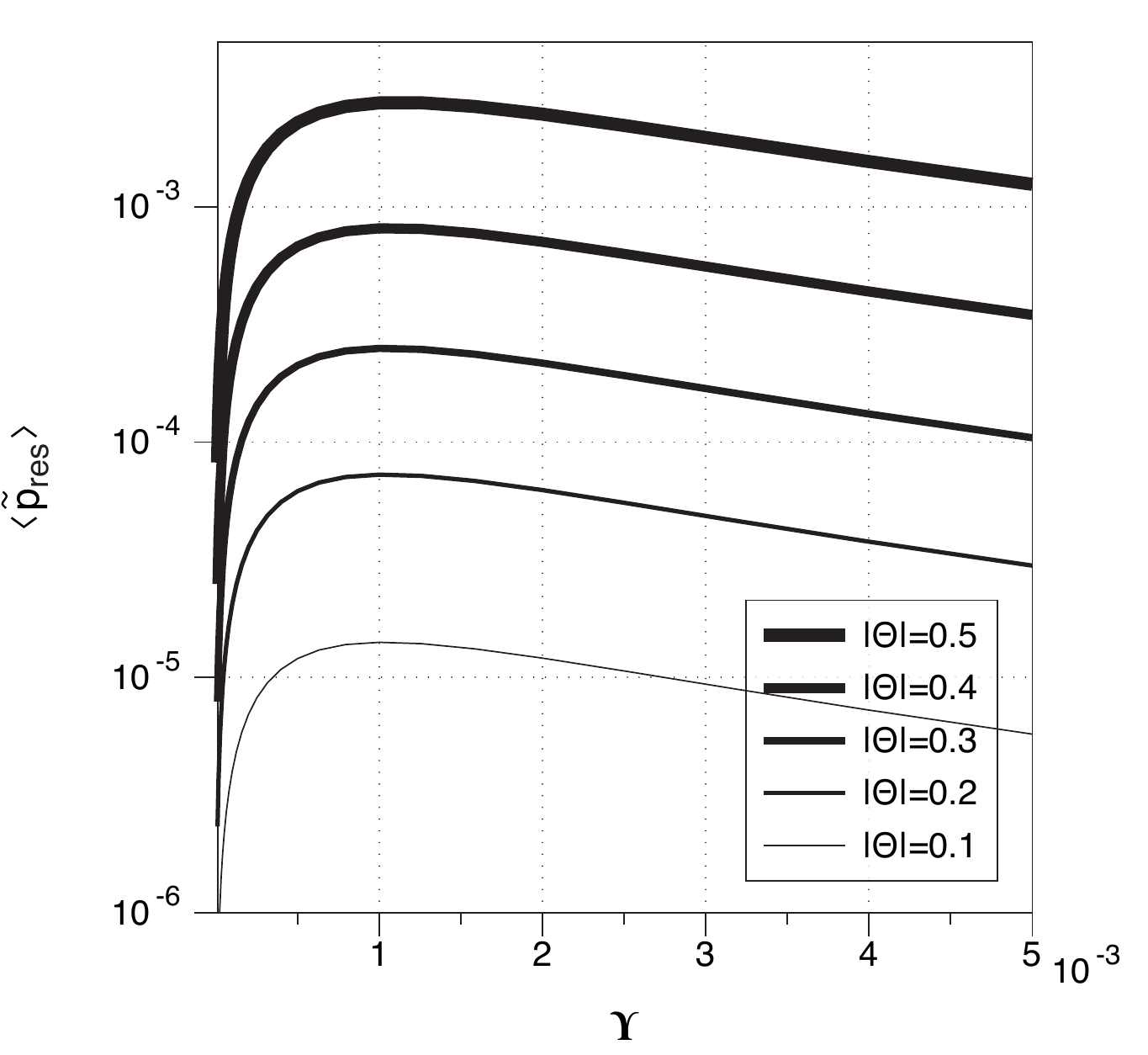}(a)
\includegraphics[trim = 0mm 0mm 0mm 0mm, clip, width=7.5cm]{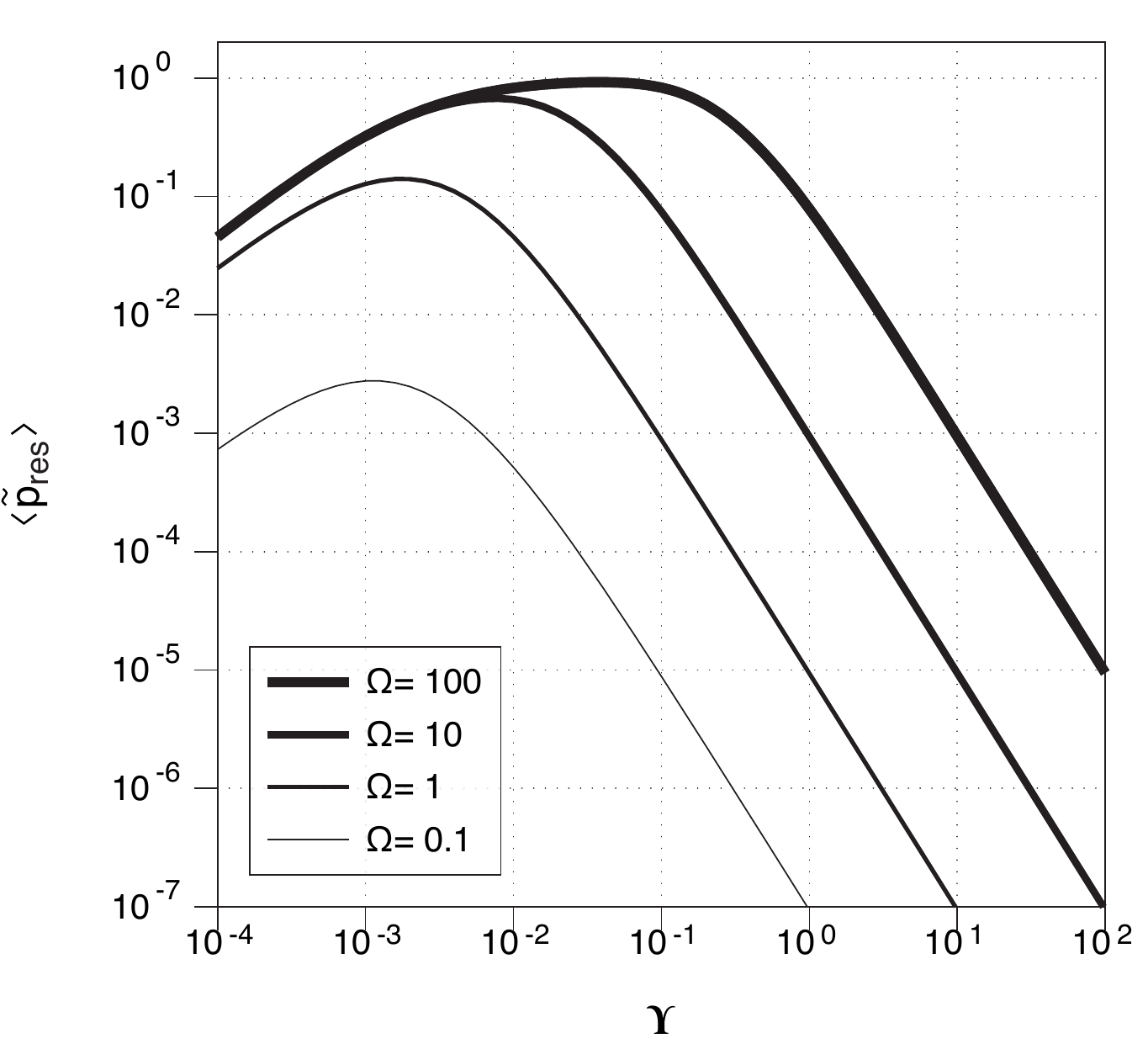}(b)\\
\includegraphics[trim = 0mm 0mm 0mm 0mm, clip, width=7.5cm]{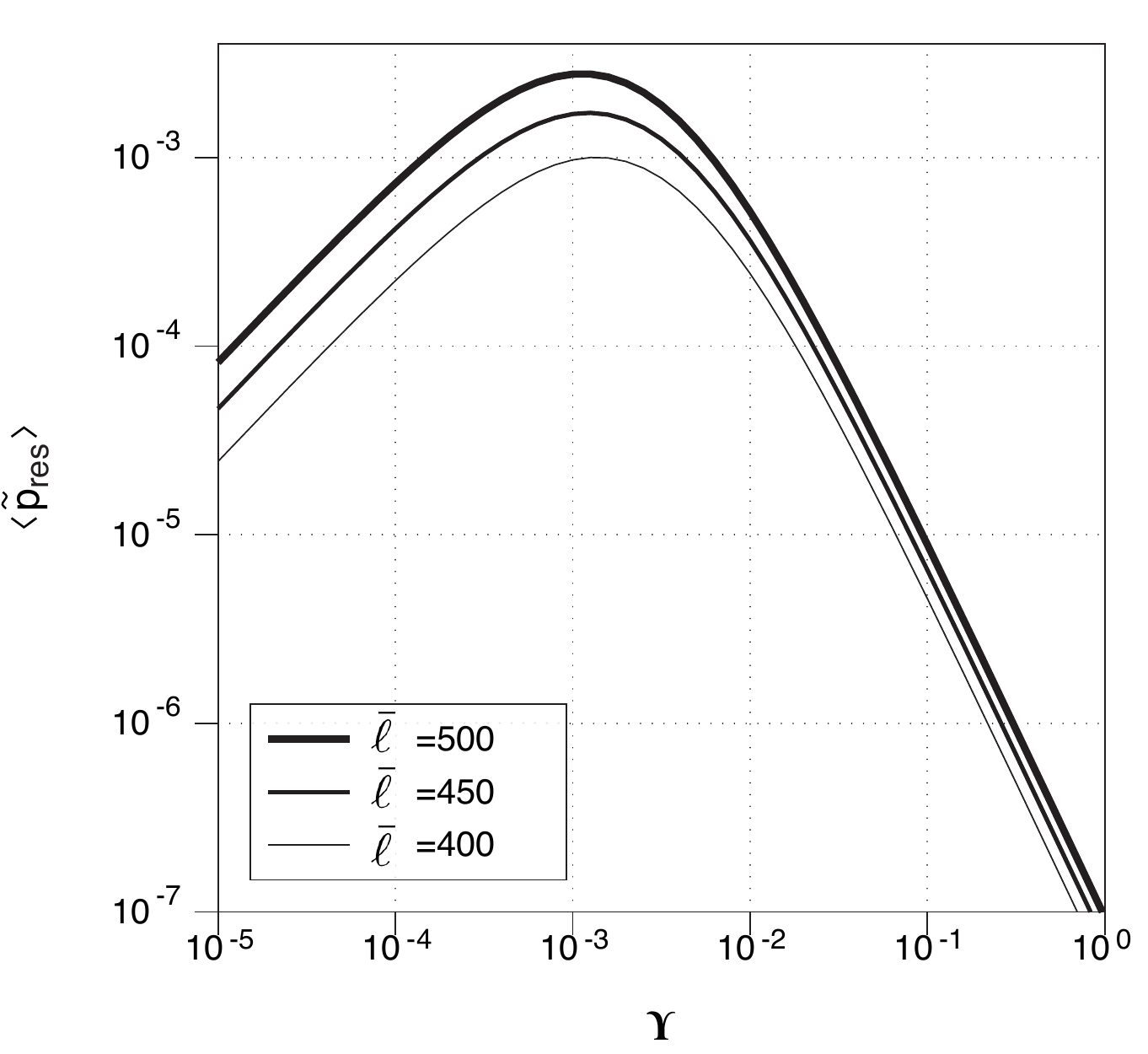}(c)
\includegraphics[trim = 0mm 0mm 0mm 0mm, clip, width=7.5cm]{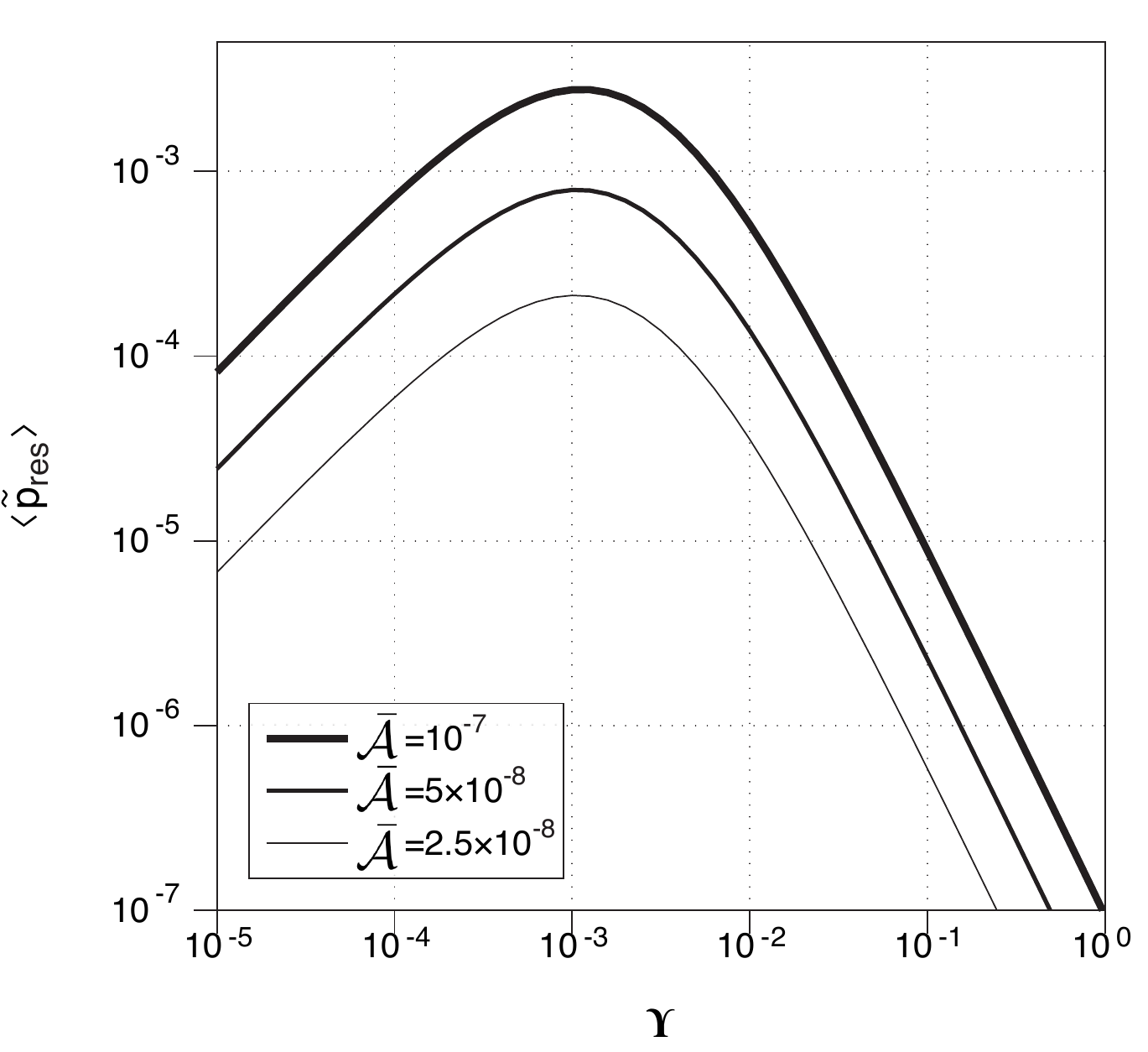}(d)
\caption{We plot the long time average reservoir pressure as a function of $\Upsilon$ for several values of $\Theta$ (a), $\Omega$ (b), $\bar{\ell}$ (c), and $\bar{\mathcal{A}}$ (d).  When the variables are not being varied, we fix $\Theta=0.5$, $\Omega=0.1$, $\bar{\mathcal{A}}=10^{-7}$, $\bar{\ell}=500$, and $\Xi=0$.}
\label{fig:presvup2}
\end{figure*}

 Qualitatively similar results were obtained for other parameters.  For example, we present the results obtained for $\Theta=0.5$, $\Omega=0.1$, $\bar{\mathcal{A}}=10^{-7}$, $\bar{\ell}=500$, and $\Xi=0$ in figure \ref{fig:presvup2}.    We have not found any set of parameters for which pumping extracts material away from the reservoir when $\Xi=0$.

\begin{figure*}
\includegraphics[trim = 0mm 0mm 0mm 0mm, clip, width=7.5cm]{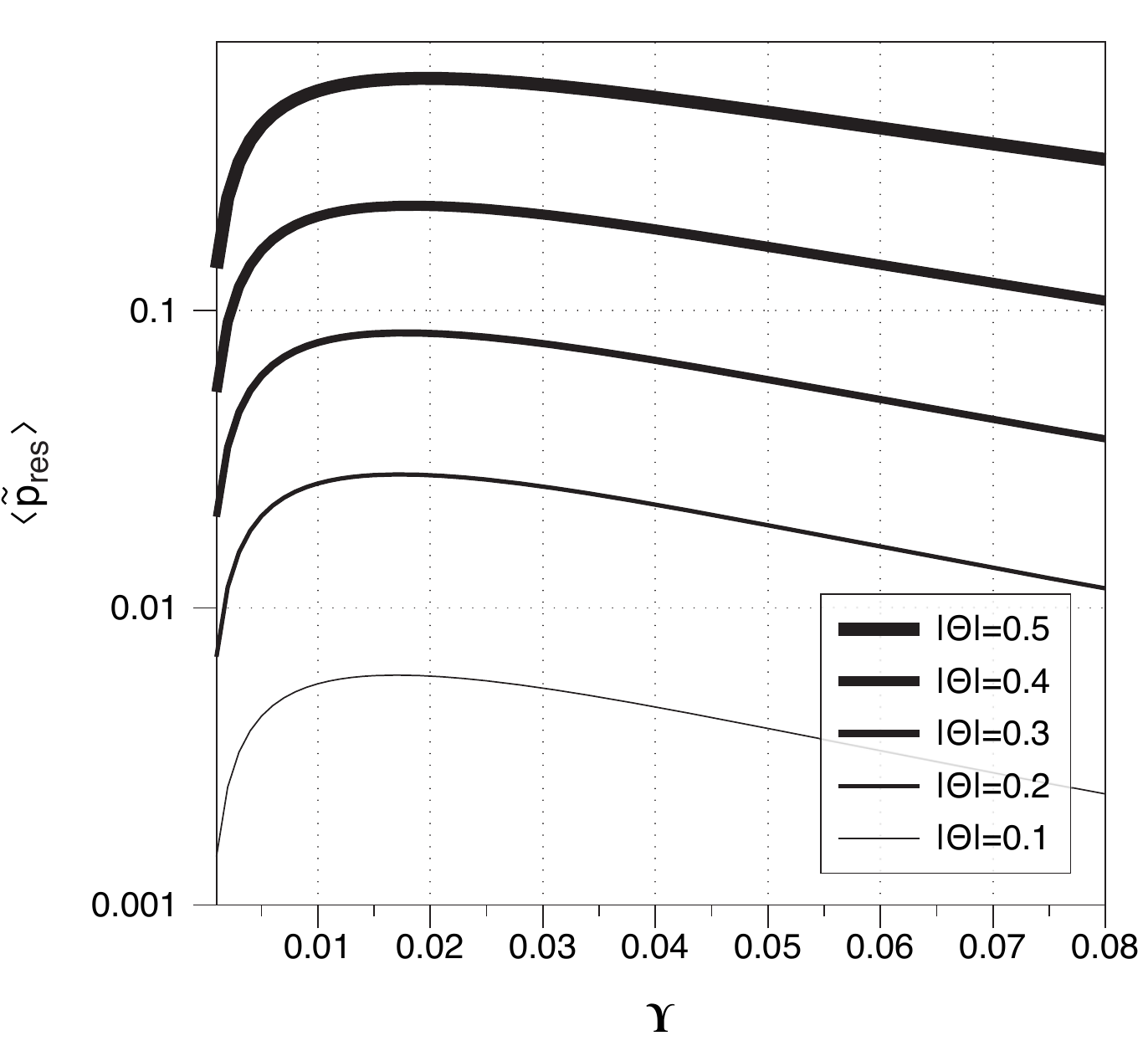}(a)
\includegraphics[trim = 0mm 0mm 0mm 0mm, clip, width=7.5cm]{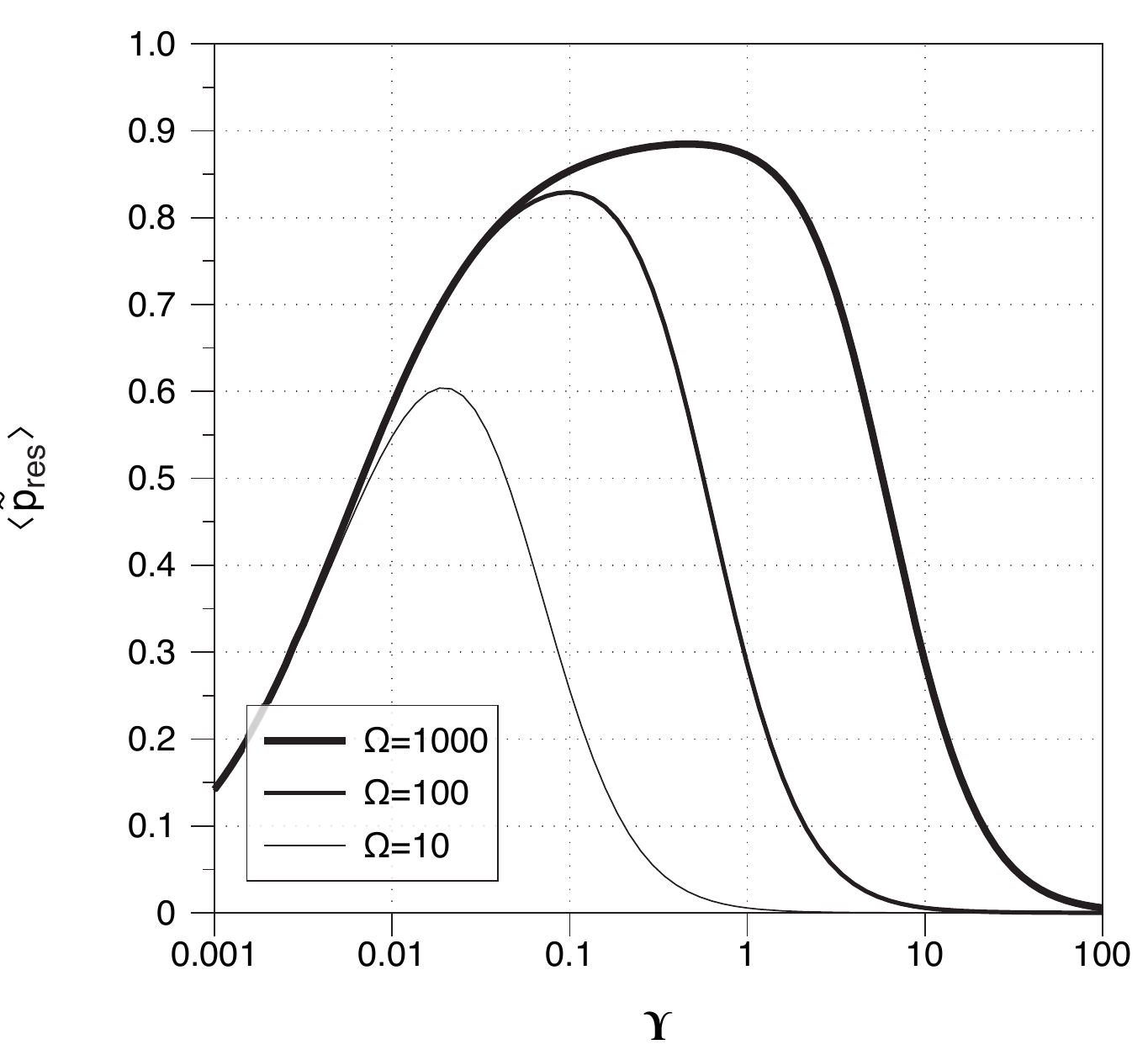}(b)\\
\includegraphics[trim = 0mm 0mm 0mm 0mm, clip, width=7.5cm]{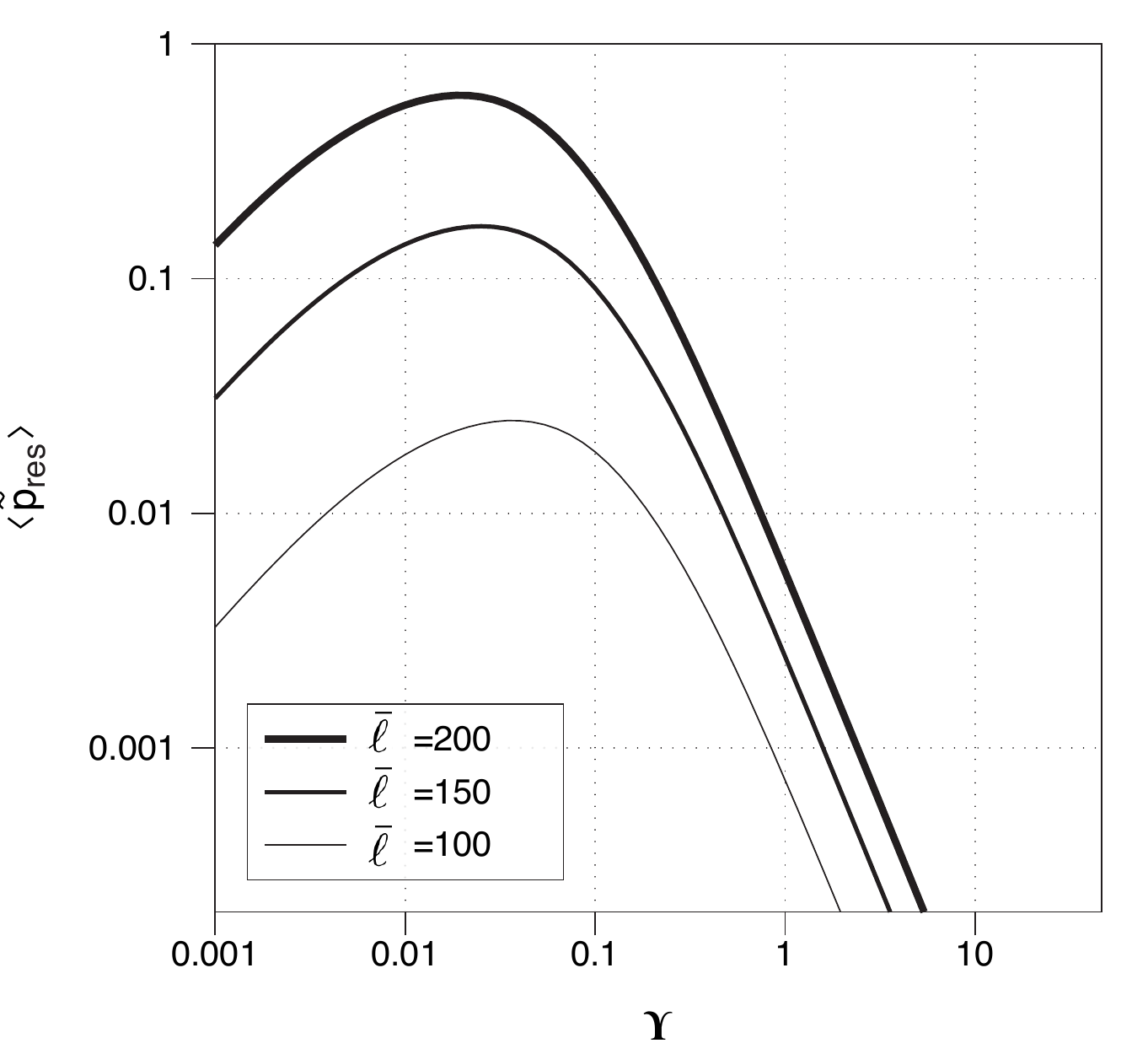}(c)
\includegraphics[trim = 0mm 0mm 0mm 0mm, clip, width=7.5cm]{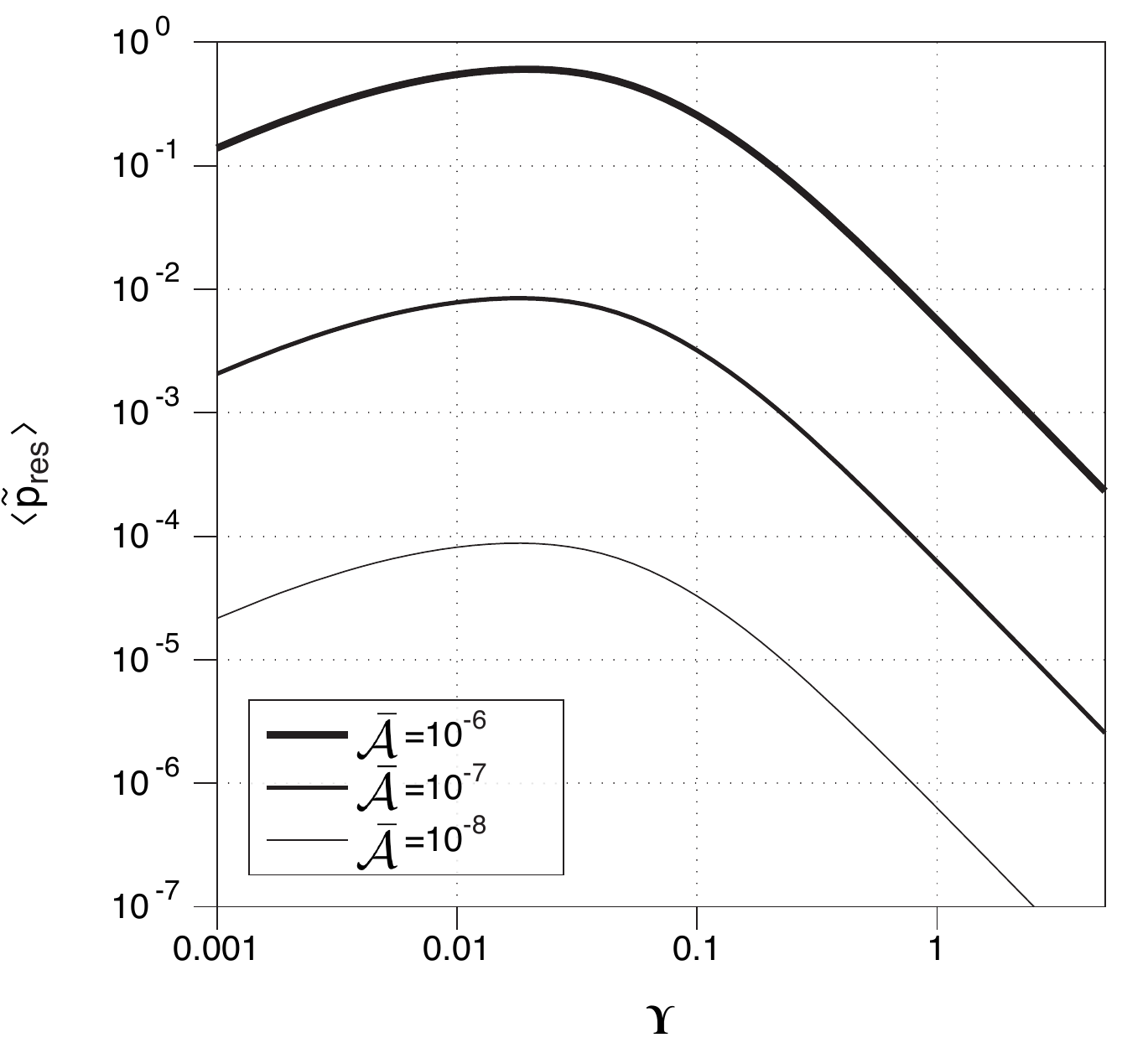}(d)
\caption{Long time average reservoir pressure as a function of $\Upsilon$ for several values of $\Theta$ (a), $\Omega$ (b), $\bar{\ell}$ (c), and $\bar{\mathcal{A}}$ (d).  By default $\Theta=0.5$, $\Omega=10$, $\bar{\mathcal{A}}=10^{-6}$, $\bar{\ell}=200$, and $\Xi=0$.}
\label{fig:presvup}
\end{figure*}

\subsection{The dependency of $\langle\tilde{p}_{res}\rangle$ on $\bar{\mathcal{A}}$, $\bar{\ell}$, and $\Omega$.}
\begin{figure}
\includegraphics[trim = 0mm 0mm 0mm 0mm, clip, width=3.5cm]{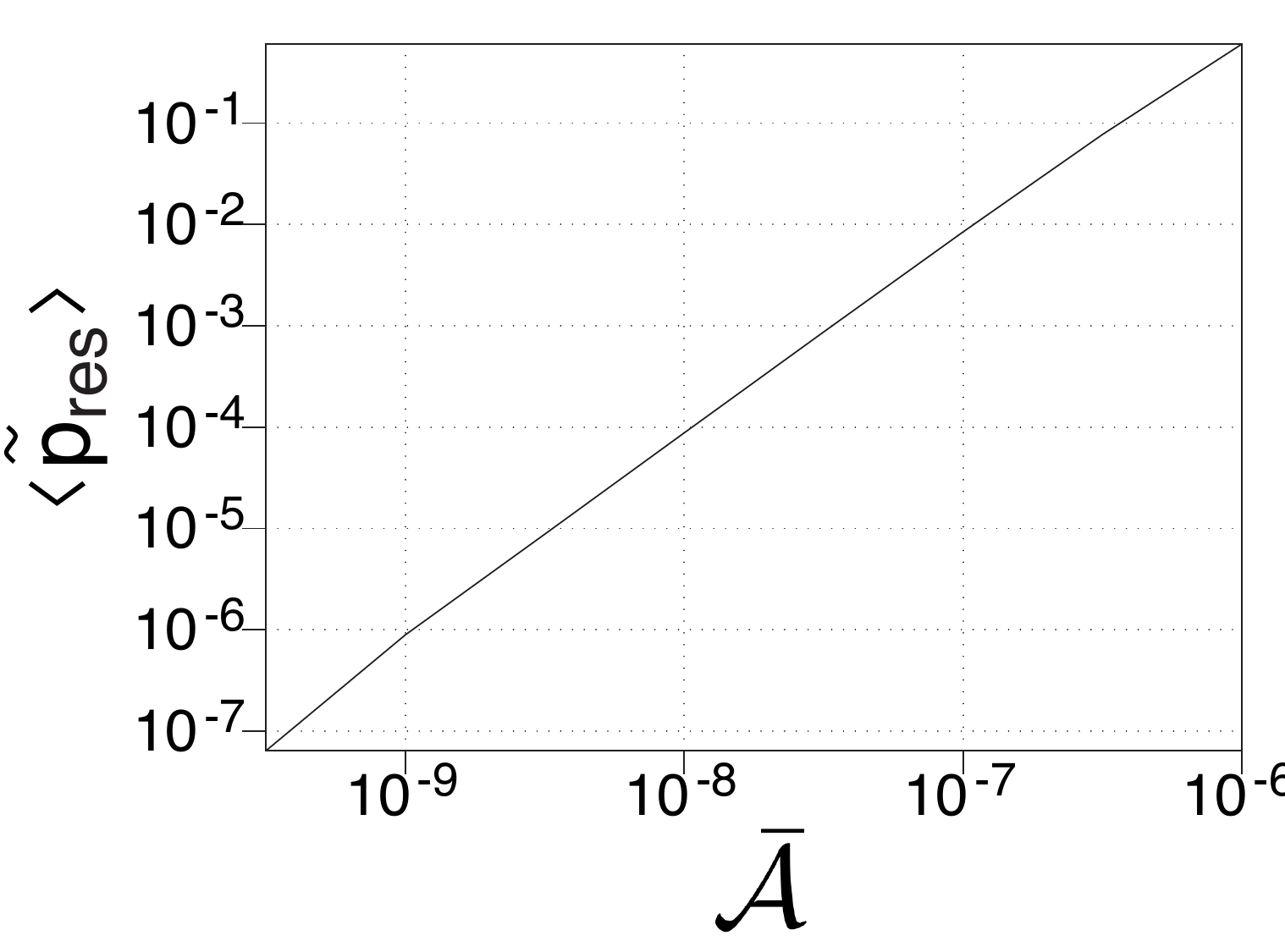}(a)
\includegraphics[trim = 0mm 0mm 0mm 0mm, clip, width=3.5cm]{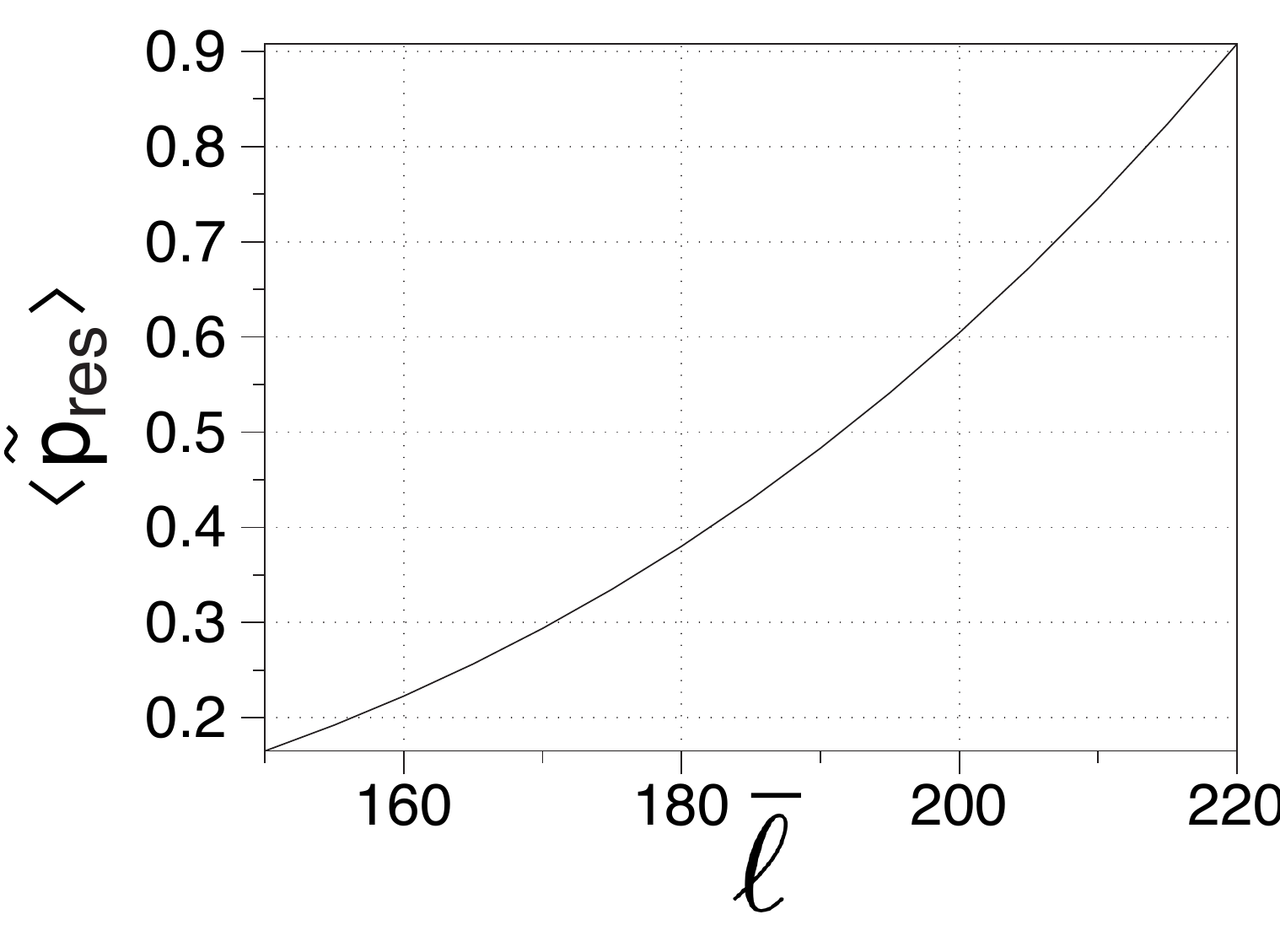}(b)
\includegraphics[trim = 0mm 0mm 0mm 0mm, clip, width=3.5cm]{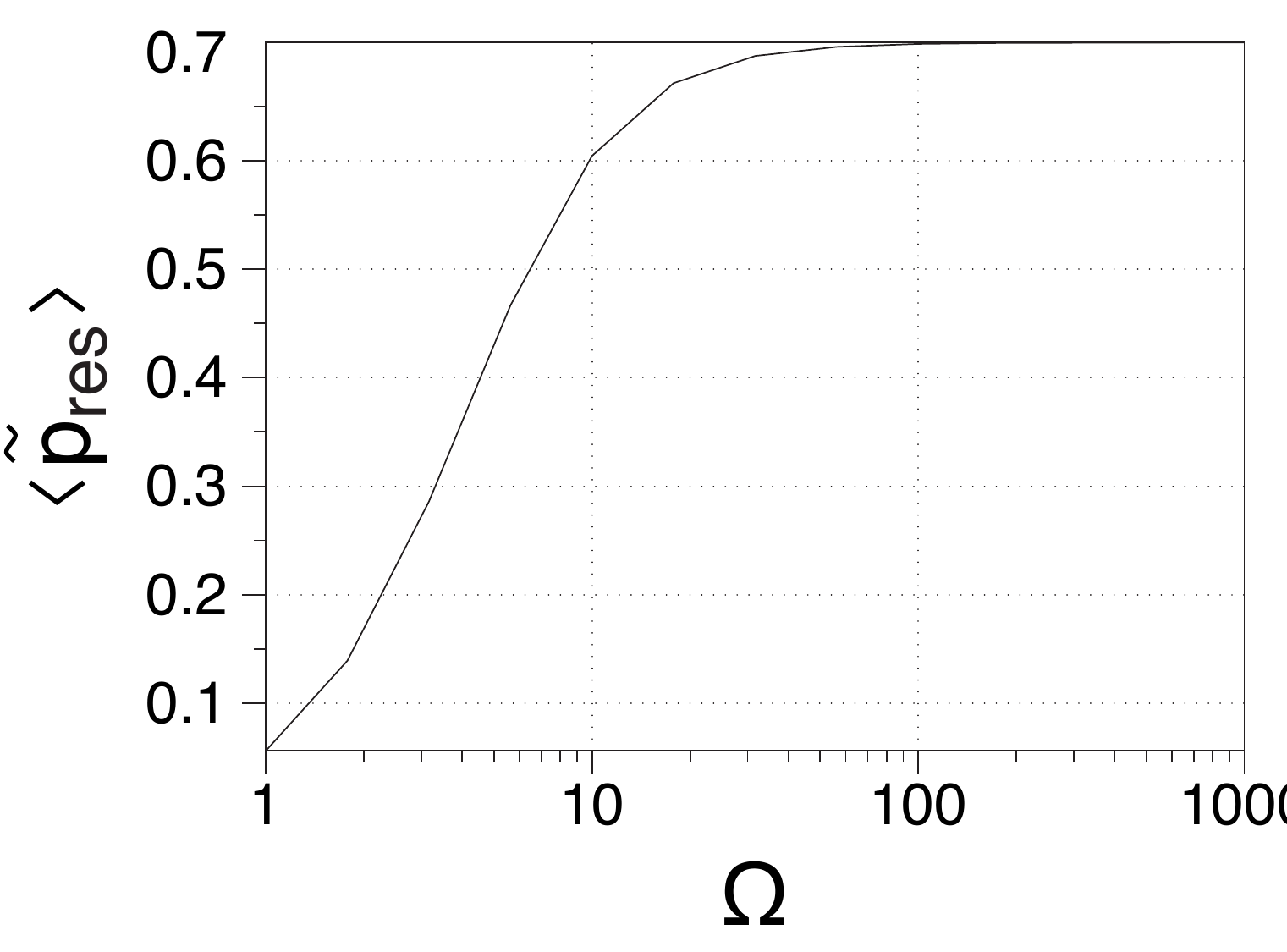}(c)
\caption{We plot the long time average reservoir pressure as a function of $\bar{\mathcal{A}}$, $\bar{\ell}$, and $\Omega$.  When not being viewed as an independent variable, $\bar{\mathcal{A}}=10^{-6}$, $\bar{\ell}=200$, and $\Omega=10$.  We also fix $\Upsilon=0.02$, $\Theta=0.3$, and $\Xi=0$.}
\label{fig:presvparam}
\end{figure}

Having determined the existence of an optimal value for $\Upsilon$ for several parameter ranges, we next inquire how changes in the other parameters alter the average amount of material extracted.  
Material extraction as a function of $\bar{\mathcal{A}}$, $\bar{\ell}$, and $\Omega$ is shown in figure \ref{fig:presvparam}.  By default $\bar{\mathcal{A}}=10^{-6}$, $\bar{\ell}=200$, $\Omega=10$, $\Upsilon=0.02$, $\Theta=0.3$, and $\Xi=0$.  Our results indicate that the mean reservoir pressure increases with $\Omega$, whereas the gain appears to be bounded from above.  Qualitatively similar results are obtained for other points in parameter space (not shown).

\begin{figure}
\includegraphics[trim = 0mm 0mm 0mm 0mm, clip, width=7cm]{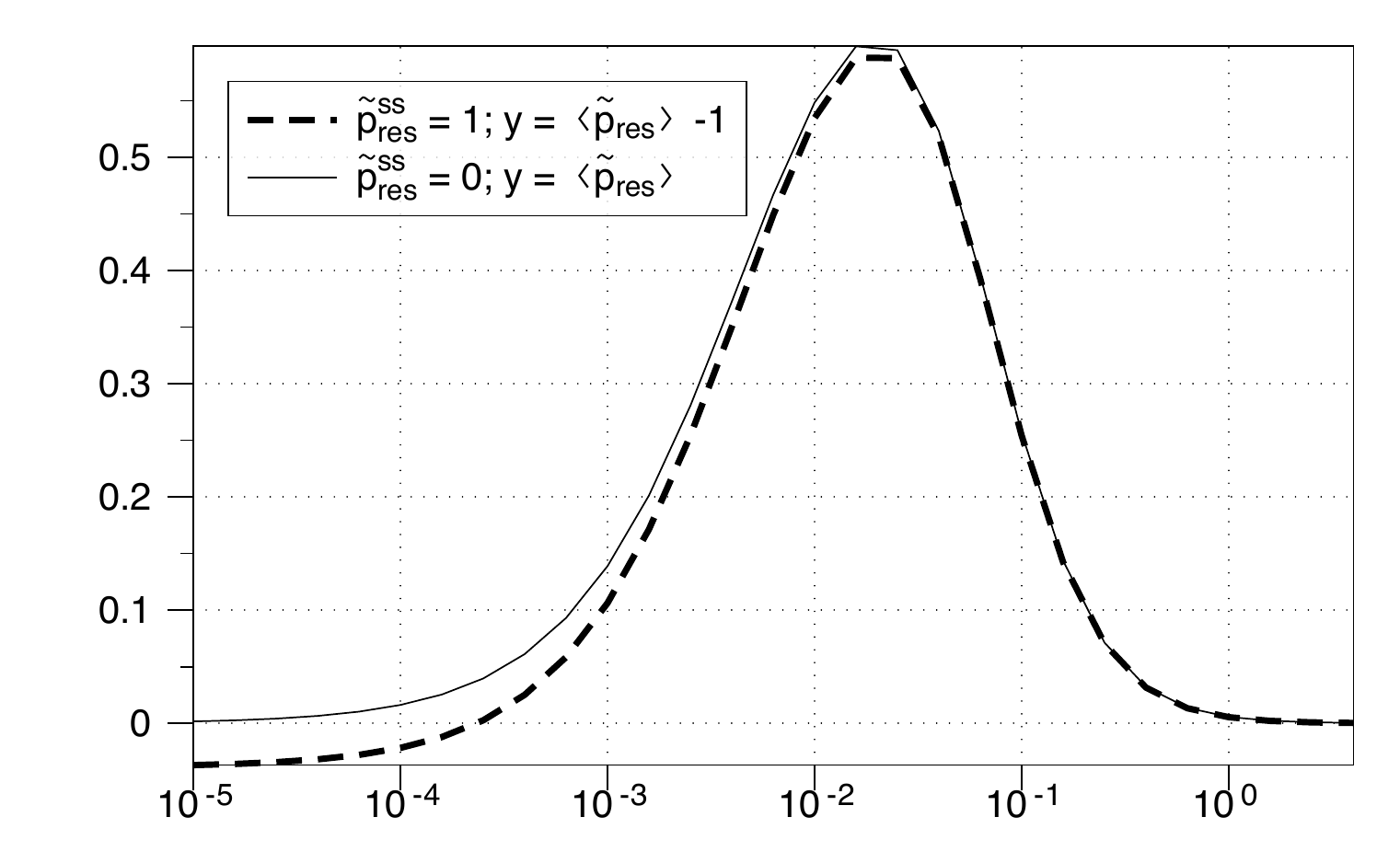}
\caption{We fix $\tilde{p}_{res}^{ss}=1$ and vary $\Upsilon$ setting $\Theta=0.5$, $\bar{\mathcal{A}}=10^{-6}$, $\bar{\ell}=200$, and $\Omega=10$.}
\label{fig:xion}
\end{figure}

%

\subsection{The effect of $\Xi$}
We conclude our parameter study by changing the average amount of material being extracted in the absence of pumping, and determine the percent gain in material extraction when pumping is introduced into the system.  Parametrically, this means that we change $\Xi$, and note that under no pumping (either $\Omega=0$ or $\Theta=0$), $\tilde{p}_{res}$ has a stable steady state value given by
\begin{align}
\tilde{p}_{res}^{ss}\equiv \lim_{t\rightarrow\infty}\tilde{p}_{res} = \frac{\Xi \bar{r}}{\lambda(0)}\times\frac{\tanh(\lambda(0)\bar{\ell})}{1+\Upsilon\tanh(\lambda(0)\bar{\ell})}.
\end{align}
We set $\bar{\mathcal{A}}=10^{-6}$, $\bar{\ell}=200$, $\Omega=10$, $\Theta=0.5$ and vary $\Upsilon$; for each value of $\Upsilon$, we set $\Xi$ so that $\tilde{p}_{res}^{ss} = 1$.  
Simulation results demonstrate that for smaller values of $\Upsilon$, material extraction is impeded by roughly 4\% (see figure \ref{fig:xion}).  To identify the mechanism of this impedance,  we first note that for $\bar{\ell}$ and $\bar{\mathcal{A}}$ fixed, the average pressure is expected to converge to $\tilde{p}_{res}^{ss} = 1$ in the limit that either $\Theta$ or $\Omega$ approach zero.  We numerically test this idea by decreasing both $\Omega$ and $\Theta$ and present our results in table \ref{table:xiinvest}.  We find that the reservoir pressure does \emph{not} approach the steady state value in the limit of $\Omega\rightarrow0$, and instead decreases until it converges to some value less than 1.  As $\Theta\rightarrow0$ however, we do see  $\tilde{p}_{res}^{ss} \rightarrow 1$.  

\begin{figure*}
\includegraphics[trim = 0mm 0mm 0mm 0mm, clip, width=5cm]{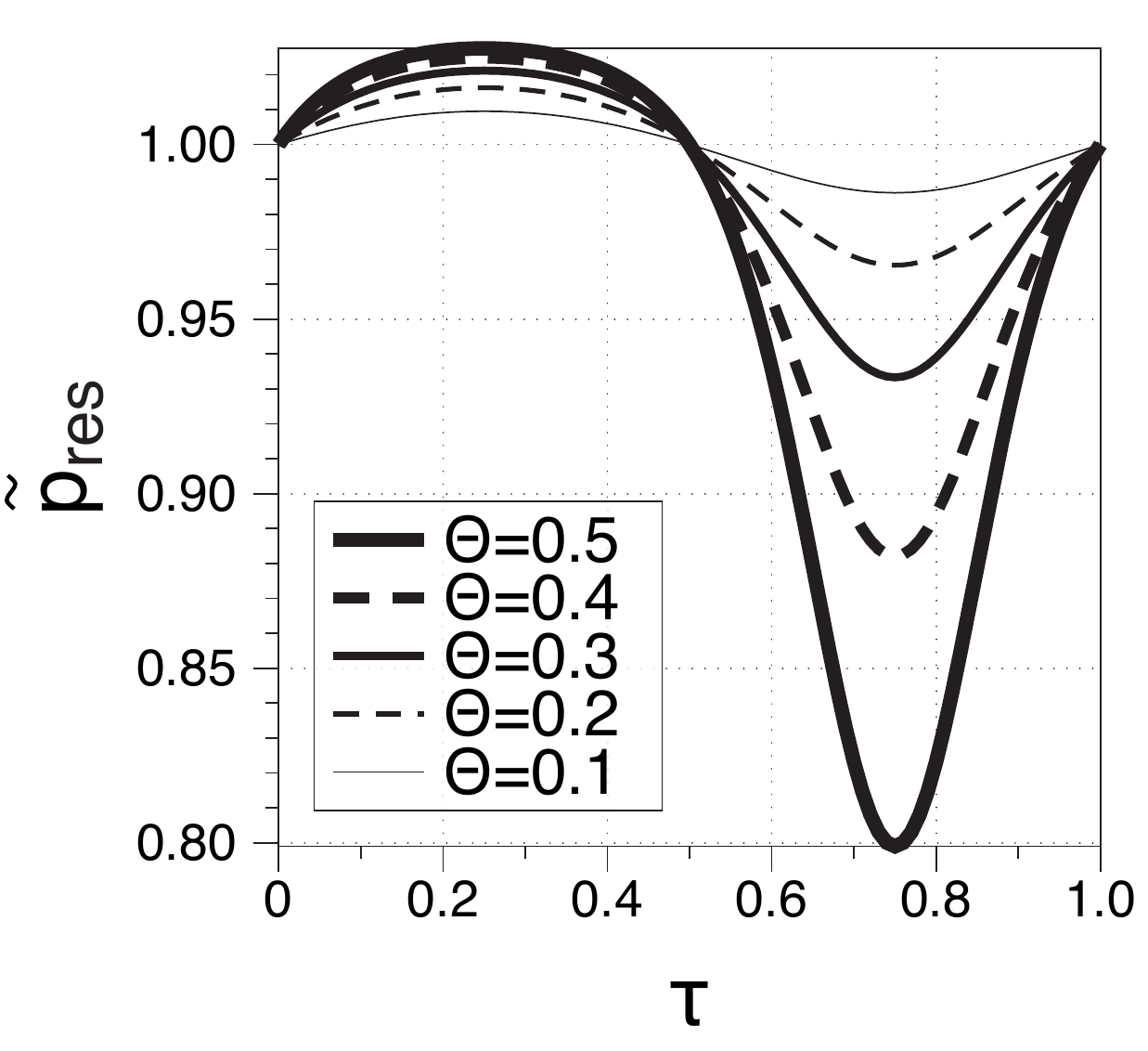}(a)
\includegraphics[trim = 0mm 0mm 0mm 0mm, clip, width=5cm]{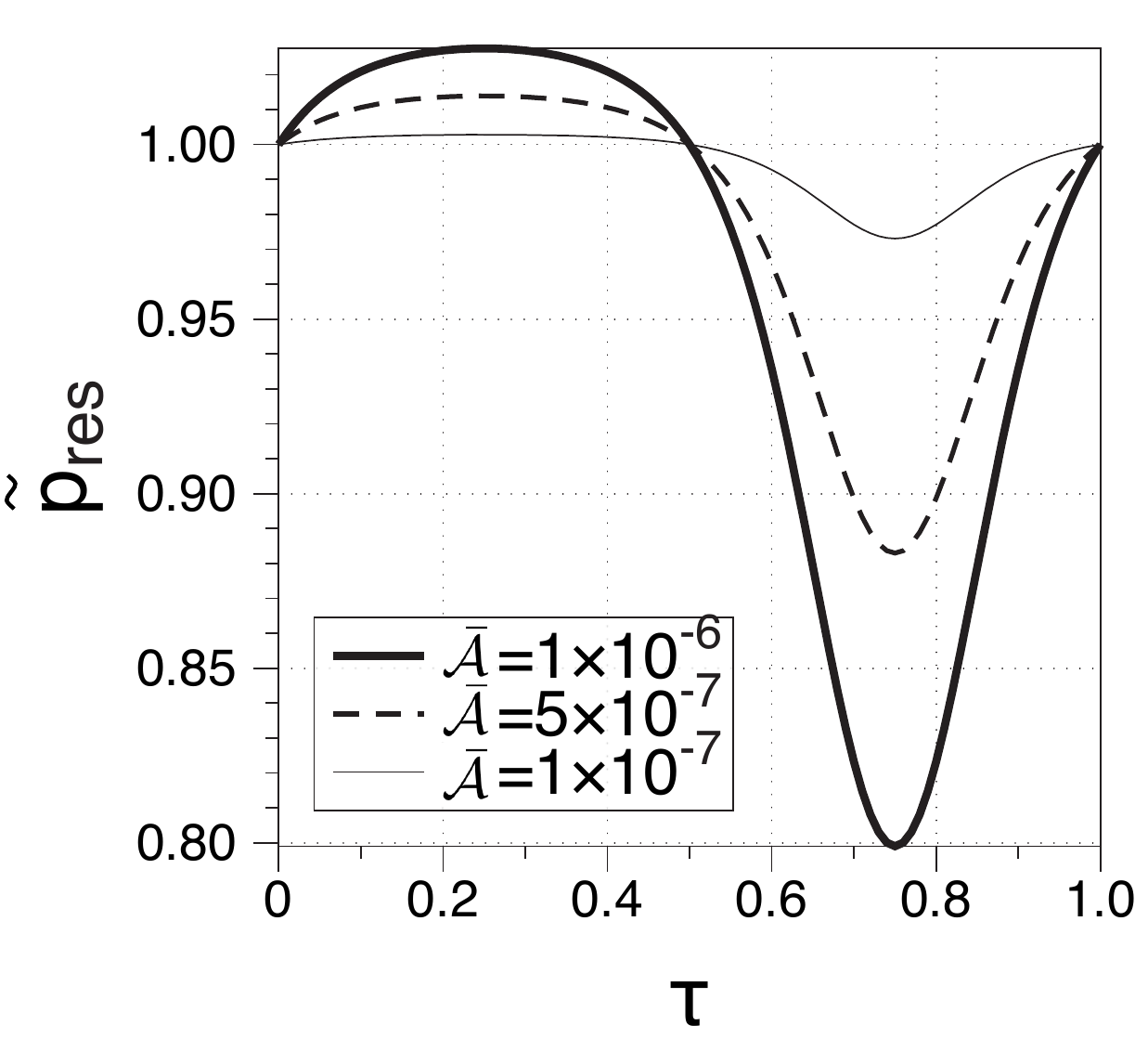}(b)
\includegraphics[trim = 0mm 0mm 0mm 0mm, clip, width=5cm]{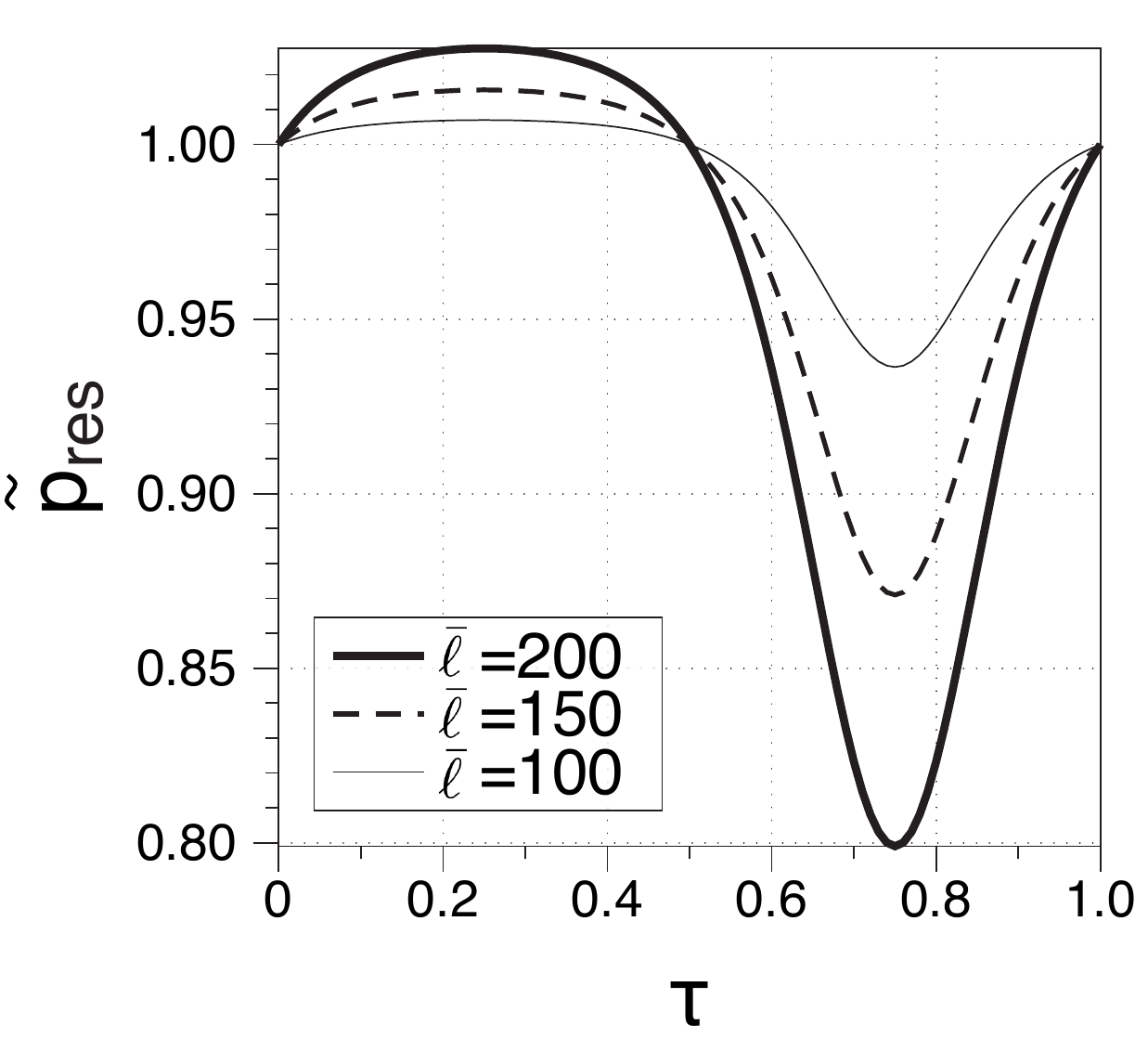}(c)
\caption{We fix $\tilde{p}_{res}^{ss}=1$ and vary $\tau$ over a full period of expansion and contraction. When not being varied, the parameters are fixed at $\Theta=0.5$, $\bar{\mathcal{A}}=10^{-6}$, and $\bar{\ell}=200$.}
\label{fig:xionpvar}
\end{figure*}

In the case that $\Omega\rightarrow0$ the contribution of the mass flux due to pumping becomes small when compared to the mass flux driven by the hydrostatic pressure.  This suggests that the third term on the righthand side of equation \ref{eqn:modelextractnondim} becomes small.  Assuming that the contribution of flux from pumping remains bounded (as it will within the radius of convergence) we can write the solution to equation \ref{eqn:modelextractnondim} with $\Xi\neq0$ as
\begin{align}
\tilde{p}_{res}(\tilde{t}) =&\tilde{p}_{res}(0) \exp\left(-\int_0^{\tilde{t}} g(s)\right)\nonumber\\
&+\Xi \exp\left(-\int_0^{\tilde{t}} g(s) \right)\ast h(\Omega \tilde{t})+O(\Omega),\label{eqn:analyticsolpumpwithXi}
\end{align}
where
\begin{align}
h(\Omega \tilde{t})\equiv\Xi\frac{g(\tilde{t})-1}{\Upsilon}.
\end{align}
To study this solution we first ignore the transient term by setting $\tilde{p}_{res}(0)=0$ and neglect the small $O(\Omega)$.  Next we set $\Upsilon=0$ to ensure that we examine the regime of parameter space in which we expect material extraction to be impeded by pumping (see figure \ref{fig:xion}).  Finally, since we will examine the small limit of $\Omega$, we set $\tilde{t} = 2\pi\tau/\Omega$ where $\tau\in[1,2]$ so that we may examine flow over a full period of pumping.  We do not choose $\tau\in[0,1]$ in order to avoid considering the transient behavior for small values of $\tilde{t}$.  We find that as $\Omega\rightarrow 0$, 
\begin{align}
\tilde{p}_{res}(2\pi\tau/\Omega) &= \int_0^{2\pi\tau/\Omega} e^{-s} h(2\pi\tau-\Omega s)ds\nonumber \\
&\sim \int_0^{2\pi\tau/\Omega} e^{-s} h(2\pi\tau)ds\nonumber\\
&= h(2\pi\tau)=\Xi\frac{g(2\pi\tau/\Omega)-1}{\Upsilon}.
\end{align}
Figure \ref{fig:xionpvar} shows $\tilde{p}_{res}$ as a function of $\tau$ for several values of $\Theta$, $\bar{\mathcal{A}}$, and $\bar{\ell}$.  We find that the reservoir pressure decreases when the channel is compressed and vice versa; however, this effect is asymmetric.  Indeed, this asymmetry enables the overall average to decrease in the mean pressure reservoir for low values of $\Omega$.  As demonstrated in figure \ref{fig:xion}, this effect is seen even when $\Omega=10$.


\begin{table}
\begin{tabular}{l | cccc}
$\Omega$  & 10 & 1 & 0.1 & 0.01\\
$\langle\tilde{p}_{res}\rangle$& 0.9631 & 0.9623 & 0.9614 & 0.9614\\
\hline
$\Theta$  & 0.4 & 0.3 & 0.2 & 0.1 \\
$\langle\tilde{p}_{res}\rangle$& 0.9791& 0.9894& 0.9956 &0.9989\\
\end{tabular}
\caption{We examine the sensitivity of $\tilde{p}_{res}^{ss}$ with respect to $\Omega$ and $\Theta$ with $\tilde{p}_{res}^{ss}=1$, $\Upsilon=10^{-5}$, $\bar{\ell}=200$, and $\bar{\mathcal{A}}=10^{-6}$.  When not being varied, $\Omega=10$ and $\Theta=0.5$.}
\label{table:xiinvest}
\end{table}

\section{Discussion}
We have developed a model for material extraction and demonstrated both qualitatively and quantitatively the mechanism by which uniform pumping enhances material extraction from a channel with permeable walls.  Furthermore, we have provided evidence of optimal conditions for material extraction in terms of the ratio between reservoir stiffness and material depletion.  To determine the flux across the channel-reservoir barrier, we have derived a formal series expansion for Stokes flow in a channel with permeable and uniformly moving walls.  Although there has been a great deal of work examining this problem (see for example \cite{Majdalani2002, Asghar2008,Asghar2010,MohyudDin2010,XinHui2011,Rashidia11,Azimi2014,Sushila2014}), we have examined a solution in which pressure gradient drives flow across the boundary rather than weak permeability.  To our knowledge, our work is the first to examine this boundary condition in the context of expanding and contracting walls.  We have investigated the convergence regimes of this series expansion and found analytic bounds for the parameter domain of convergence, and have numerically determined the convergence regimes with respect to the nondimensional permeability and channel length.  

The simplicity and limitations of the present model when applied to biological processes such as insect respiration or renal filtration must be acknowledged.  One discrepancy is that while the present model represents uniformly moving walls, compression of the biological channel typically occurs at a discrete location (insect respiration\cite{Abo2011}) or in a wave like pattern (renal filtration\cite{SchmidtNielsen2011}).  Secondly, exchange across the channel-reservoir barrier is typically not as simple as a single homogenized fluid.  In insect respiration, air may be thought of as a mixture of oxygen and unusable elements such as nitrogen and carbon dioxide.  To model this distinction a phase field model may be more suitable for the air flow, and in addition only oxygen should be depleted (and converted into waste) within the reservoir.   However, if the permeabilities for the waste and oxygen are equivalent, then the phase field model would decouple from the flow equations, and the work presented here would be a first step in solving such a model.  In renal filtration, not only water is being filtered, but ions as well.  Thus osmotic pressures must be taken into account when considering fluid exchange across the channel-reservoir barrier.  Despite these limitations, having obtained a promising first result on the effect of pumping on fluid extraction, we plan to investigate these more realistic models in future work.

\section*{Acknowledgements} 
This research was supported by the National Science Foundation (NSF) grant DMS1263995, to A. Layton; Research Training Groups grant DMS0943760 to the Mathematics Department at Duke University; KI-Net NSF Research Network in the Mathematical Sciences grant No. 1107291, and NSF grant DMS1514826 to Jian-Guo Liu.  

%

\end{document}